\documentclass[twocolumn]{revtex4-2}
\usepackage{xcolor}
\usepackage{tikz}
\usepackage{bbding}
\usepackage{mathtools}
\usepackage{amssymb}
\usepackage{graphicx}
\usepackage{dcolumn}
\usepackage{bm}
\usepackage{hyperref}

\usepackage{url}

\usepackage{academicons}
\newcommand{\orcid}[1]{\href{https://orcid.org/#1}{\textcolor[HTML]{A6CE39}{\aiOrcid}}}

\begin{document}

\preprint{APS/123-QED}


\title{Transfer of quantum states and stationary quantum correlations in a hybrid optomechanical network}

\author{Hugo Molinares $^{1}$}
\author{Bing He $^{2}$}
\email{bing.he@umayor.cl}
\author{Vitalie Eremeev $^{3,4}$}
\email{corresponding author: vitalie.eremeev@udp.cl}

\affiliation{
$^{1}$Departamento de Ciencias Físicas, Universidad de La Frontera, Casilla 54-D, Temuco 4780000, Chile
}
\affiliation{
$^{2}$Centro de Optica e Informaci\'on Cu\'antica, Universidad Mayor,\\
camino la Piramide 5750, Huechuraba, Santiago, Chile
}
\affiliation{
$^{3}$Instituto de Ciencias B\'asicas, Facultad de Ingenier\'ia y Ciencias, Universidad Diego Portales, Av.
 Ejercito 441, Santiago, Chile
}
\affiliation{
$^{4}$Institute of Applied Physics, Academiei 5, MD-2028, Chi\c{s}in\u{a}u, Moldova
}

\date{\today}

\begin{abstract}
We present a systematic study on the effects of dynamical transfer and steady-state synchronization of quantum states in a hybrid optomechanical network, consisting of two cavities with atoms inside and interacting via a common moving mirror (i.e. mechanical oscillator), are studied. It is found that high fidelity transfer of Schr\"{o}dinger's cat and squeezed states between the cavities modes is possible. Additionally, we show the effect of synchronization of cavity modes in a steady squeezed states at high fidelity realizable by the mechanical oscillator which intermediates the generation, transfer and stabilization of the squeezing. In this framework, we also have studied the generation and evolution of bipartite and tripartite entanglement and found its interconnection to the effects of transfer and synchronization. Particularly, when the transfer occurs at the maximal fidelity, at this instant any entanglement is almost zero, so the modes are disentangled. On the other hand, when the two bosonic modes are synchronized in a squeezed stationary state, then these modes are also entangled. The results found in this study may find their applicability in quantum information and computation technologies, as well in metrology setups, where the squeezed states are essential.
\end{abstract}

\maketitle

\section{Introduction}\label{sec1}
The protocols of generation, protection and transfer of coherent and non-classical states in quantum systems are considered of great importance in the era of the third quantum revolution \cite{Celi2016}. For example, squeezed states of light, spins or mechanical oscillator (MO) motion are particular ingredients of many applications in metrology \cite{AVS2019,AVS2020, Quantum2020, APL2022}, sensing \cite{And2017,Mas2019}, e.g. gravitational-wave detection \cite{LIGO2013}, continuous-variable information processing \cite{RMP2005,PRAp2019}, etc. 

Nowadays, squeezing is often produced and controlled in spin/opto-mechanical systems, generally known as hybrid systems and considered very useful for generating and transferring the squeezing of different degrees of freedom such as photons, phonons and spins \cite{Pur2013,Pir2015, OL2018,PRA2022,OE2019}. Furthermore, to build many quantum architectures, it is essential to engineer effective protocols for the ignition and transfer of the nonclassical states between the separate systems which integrate a quantum network. There are several studies proposing protocols of quantum state transfer in optomechanical configurations \cite{Clerk2012,Wal2010,Bai2019}. As an alternative to the transfer protocol, one can expect of performing a quantum dynamical synchronization of the degrees of freedom in the hybrid system, and even for the entire network in some particular cases. Synchronization here refers to those between the dynamical stabilization of a target steady-state for several degrees of freedom, e.g. \cite{Bra2021,PRA2022}. Generally, such protocol is not trivial to realize in the quantum systems \cite{Jak2022}. Moreover, for the real systems, it is crucial to protect the target quantum states from the intrinsic losses, decoherence effects.

A kind of quantum stable synchronization protocol we propose in the present work, i.e. a synchronization of the photonic and phononic modes in a high-fidelity squeezed states for an open optomechanical system assisted by driven three-level atoms. We show that such a synchronization can be realized by a coherent pump of squeezed phonons or photons in order to initialize squeezing in one of the modes, i.e. MO or first cavity. The mechanism of the squeezing initialization using by us is inspired by the milestone work of Walls \cite{Walls}. Additionally to synchronization of the stationary squeezed states of a pair of modes it is of particular interest to study the quantum correlations between these modes. About two decades ago some studies have shown how the multipartite entanglement within the continuous variables is related to a joint quadrature squeezing \cite{Loock99, Jing2003, Fur2003, Yone2004}. For example, in a recent work \cite{He2023} the authors propose an experimental verification of the quadripartite entanglement by measuring squeezing in a joint amplitude and phase quadratures. In this context, it is appealing to study and understand the correlation effect between the multipartite entanglement and squeezing in optomechanical systems.

We highlight the increasing interest in the systems of optomechanical networks due to their wide spectrum of applications \cite{Aspelmeyer,Luk2010, Dant2012,Wan2015}. For example, the double-cavity, (known also as mirror-in-the-middle) optomechanical system, which can be considered as simplest type of network, were studied in various proposals as \cite{Marq2012, arx4067, Bing2017, OE2019, Bur2020, Orsz2022}. In the present work we will consider a similar kind of double-cavity optomechanical system, where the middle mirror is movable (i.e. mechanical oscillator) and highly reflective on both sides, there is no transmission through the mirror. The full system in our model contains a three-level atom placed on each side of the movable mirror. The hybrid optomechanical systems carrying three-level atoms were highly interested for their important role for different protocols such as entanglement formation, mechanical cooling, sensing, etc. For example, in \cite{Ghasemi:19} a quantum repeater protocol for distributing the entanglement between two distant three-level atoms using an arrangement of QED-optomechanical hybrid was proposed. Another work \cite{Li_2018}, considers a hybrid optomechanical cooling with a three-level atomic ensemble fixed in a strong excited optical cavity. A scheme in \cite{Zhou11} proposes to apply three-level cascade atoms to entangle two optomechanical oscillators as well as two-mode fields. In the present work, the role of driven three-level atoms and MO is to stimulate and control the quantum protocols as the state transfer and stable synchronization of the squeezing in photonic and phononic modes. Taking into account the results of high-efficiency state transfer and squeezing synchronization in the simplest model of a two-mode double-cavity optomechanics, it is possible to extrapolate the model to a network with several mechanical oscillators and many cavity modes. This kind of optomechanical network is expected to be efficient for the applications in quantum technologies as sensing, metrology, transmission of quantum states and correlations, etc.

This work is organized as follows. In Sec. \ref{sec2} we present the conceptual model of the hybrid optomechanical network and a brief analysis of the role of the driving fields that stimulate the three-level atoms. In Sec. \ref{sec3} we evaluate the dynamical transfer of the quantum states between the cavities. Here, using two different quantum states as initial condition in the first cavity, we show how these states are transferred to the second cavity. Next, in Sec. \ref{sec4} we show how the bipartite and tripartite correlations are generated and evolved for the lossless dynamics. We analyze the dynamical effect of the correlations as compared to the transfer of quantum states between the cavities. Sec. \ref{sec5} is devoted to the study of the stationary synchronization of squeezing between two bosonic modes as function of the optomechanical and Jaynes-Cummings couplings. Moreover, a comprehensive analysis of the entanglement generation as a consequence of the squeezing synchronization effect is presented. Finally, we discuss and conclude our findings in Secs.\ref{sec6}-\ref{sec7}.

\begin{figure*}[t]
\centering
\includegraphics[width=0.7\linewidth]{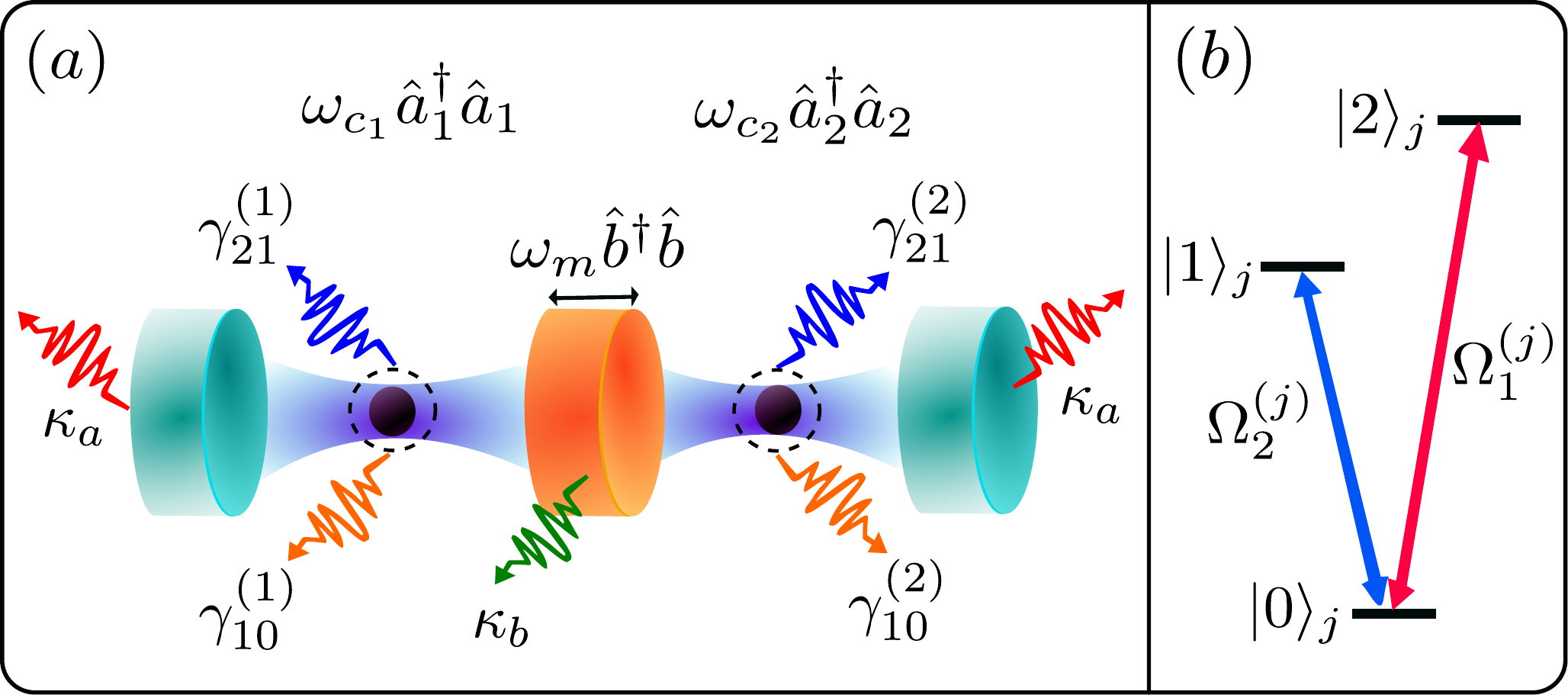}
\caption{$(a)$ Schematic diagram of a cavity-atom-mechanics system. $(b)$ Two lasers with intensities $\propto$ to $\Omega^{(j)}_{1}$ and $\Omega^{(j)}_{2}$ driving the three-level $j$-atom, which are resonant with the transitions of the levels $|2\rangle_{j} \longleftrightarrow |0\rangle_{j}$ and $|1\rangle_{j} \longleftrightarrow |0\rangle_{j}$, respectively.}
\label{fig1}
\end{figure*}

\section{Hybrid optomechanical system}\label{sec2}

Our concerned hybrid atom-cavity-mechanics system is illustrated in Fig. \ref{fig1}. The hybrid system is composed of two optical cavities coupled to a joint mechanical oscillator (MO) through nonlinear optomechanical interaction, similar to setups as \cite{Bing2017, OE2019}, besides each cavity is coupled to the upper two levels of a three-level atom. The total system's Hamiltonian is (with $\hbar=1$)
\begin{eqnarray}\label{base} 
\mathcal{H}&=&\omega_{m}b^{\dagger}b + \sum_{j=1}^{2} \bigg[ \omega_{c_{j}}a_{j}^{\dagger}a_{j} + \sum_{i=0}^{2}\omega_{i,j}\sigma_{ii,j}  \nonumber \\
&+& g_{j}\big(a_{j}\sigma^{+}_{21,j}+a_{j}^{\dagger}\sigma^{-}_{21,j}\big)+\left(-1\right)^{j}\lambda a_{j}^{\dagger}a_{j}\left(b+b^{\dagger}\right) \bigg ],
\end{eqnarray}
where $\omega_{i,j}$ are the energy levels of the three-level $j$-atom, $a_{j}(a_{j}^{\dagger})$ and $b(b^{\dagger})$ are the annihilation (creation) operators of the $j$-cavity and the MO, respectively. The Jaynes-Cummings type interaction between the two upper levels of the three-level $j$-atom and the $j$-mode of the cavity field of frequency $\omega_{c_{j}}$ is quantified by the coupling constant $g_{j}$. The interaction between the cavities and the MO of frequency $\omega_{m}$ corresponds to the standard optomechanical coupling and is quantified by the coupling constant $\lambda$. The atomic operators of lowering (raising) denoted as $\sigma^{-}_{kl,j}(\sigma^{+}_{kl,j})=|k\rangle_{j}\langle l|(|l\rangle_{j}\langle k|)$ obey standard anti-commutation relations.

\subsection{Effective atom-photon-phonon interaction}\label{sec:level1b}
An essential step to see the mutual couplings between the different elements in the hybrid system is to derive the following effective Hamiltonian in an interaction picture (see the details in Appendix \ref{apendice1}):
\begin{eqnarray}\label{h1}
\mathcal{H}_{1}=\sum_{j=1}^{2}g_{j}a_{j}\sigma^{+}_{21,j}\exp{\left\{\imath\left(\Delta_{j}t-(-1)^{j}F(t)\right)\right\}} + H.c.
\end{eqnarray}
Here we have used the Hermitian operator $F(t)\equiv\frac{\lambda}{\imath \omega_{m}}\left(b^{\dagger}\eta+b\eta^{*}\right)$, with $\eta\equiv e^{\imath\omega_{m}t}-1$, and $\Delta_{j}\equiv\omega_{2,j}-\omega_{1,j}-\omega_{c_{j}}$ is the detuning. 

In what follows, we detail the conditions and parameter regimes where the studied effects are possible. We consider the blue-detuned regime $\Delta_{j}=-\omega_{m}$ that selects two possible processes: $i)$ a photon is created, a phonon is absorbed and the atom decays from state $|2\rangle_{j}$ to $|1\rangle_{j}$, and $ii)$ a photon is absorbed, a phonon is created and the atom is excited from state $|1\rangle_{j}$ to $|2\rangle_{j}$. 

According to current experimental results, the optomechanical coupling cover a wide spectrum of values \cite{Aspelmeyer}. Therefore, in Eq. \ref{h1} we can neglect the fast oscillations of the mechanical frequency in the weak coupling regime, $\lambda\ll\omega_{m}$ \cite{Murch,Painter,Xuereb}. Considering the aforementioned, the above Hamiltonian becomes (for more details see Eqs.\ref{eqA11}-\ref{eqA13} in Appendix \ref{apendice1})
\begin{eqnarray} \label{h2}
    \mathcal{H}_{2}=\sum_{j=1}^2 (-1)^{j+1}\Lambda_{j}a_{j}\sigma^{+}_{21,j}b^{\dagger}+ H.c.,
\end{eqnarray}
where $\Lambda_{j}\equiv g_{j}\cdot \lambda/ \omega_{m}$ is the tripartite atom-photon-phonon interaction strength.

We would highlight that the Hamiltonian in Eq.\ref{h2} quantifies the effective tripartite interaction in our hybrid system and plays the key role to realize the protocols of quantum state transfer and squeezing synchronization, which will be studied in the following sections. It enables the quantum states in the first cavity to be transferred to the MO, later from MO to the second cavity and vice-versa for the unitary dynamics, i.e a reversible process. The role of the tripartite coupling in the squeezing transfer process is described in detail in Sec.\ref{sec6}A. 
In case of dissipative dynamics there is only possible the irreversible process, finally evolving the cavities and MO to stationary states, as observed in Figs. \ref{fig5}-\ref{fig6}. 

\subsection{\label{sec:level1b}Atomic driving}

In order to realize the effects of transfer and synchronisation of the quantum states between the two cavities, we consider atomic pump mechanisms by using lasers of strengths $\Omega^{(j)}_{1}$ and $\Omega^{(j)}_{2}$ for the atom $j=\{1,2\}$, respectively. As shown in Fig. \ref{fig1}($b$), these lasers are resonant with the transitions $|2\rangle_{j} \longleftrightarrow |0\rangle_{j}$ and $|1\rangle_{j} \longleftrightarrow |0\rangle_{j}$, respectively. These coherent drives are described by the Hamiltonian in the interaction picture
\begin{eqnarray}
    \mathcal{H}_{L}=\sum_{j=1}^{2} \Omega_{1}^{(j)}\left(\sigma_{20,j}^{-}+\sigma_{20,j}^{+}\right)+ \Omega_{2}^{(j)}\left(\sigma_{10,j}^{-}+\sigma_{10,j}^{+}\right).
\end{eqnarray}
Assuming that the $|1\rangle_{j} \longleftrightarrow |0\rangle_{j}$ transition (coupled to the classical field $\Omega^{(j)}_{2}$) and the $|2\rangle_{j} \longleftrightarrow |1\rangle_{j}$ transition (coupled to the quantum cavity field) are dipole allowed, i.e. involving states of opposite parity, then the driving field $\Omega^{(j)}_{1}$ for the $|2\rangle_{j} \longleftrightarrow |0\rangle_{j}$ transition, will couple states with the same parity, thus dipole forbidden. To achieve this coupling, we can use a non-linear process as an effective coherent pump from a Raman-like configuration resonant to the carrier transition where a fourth level was present and adiabatically eliminated. Moreover, for a better understanding of the role of driving fields, we discuss this in the Sec.\ref{sec6}A.

\begin{figure*}[t]
\centering
\includegraphics[width=0.32\linewidth]{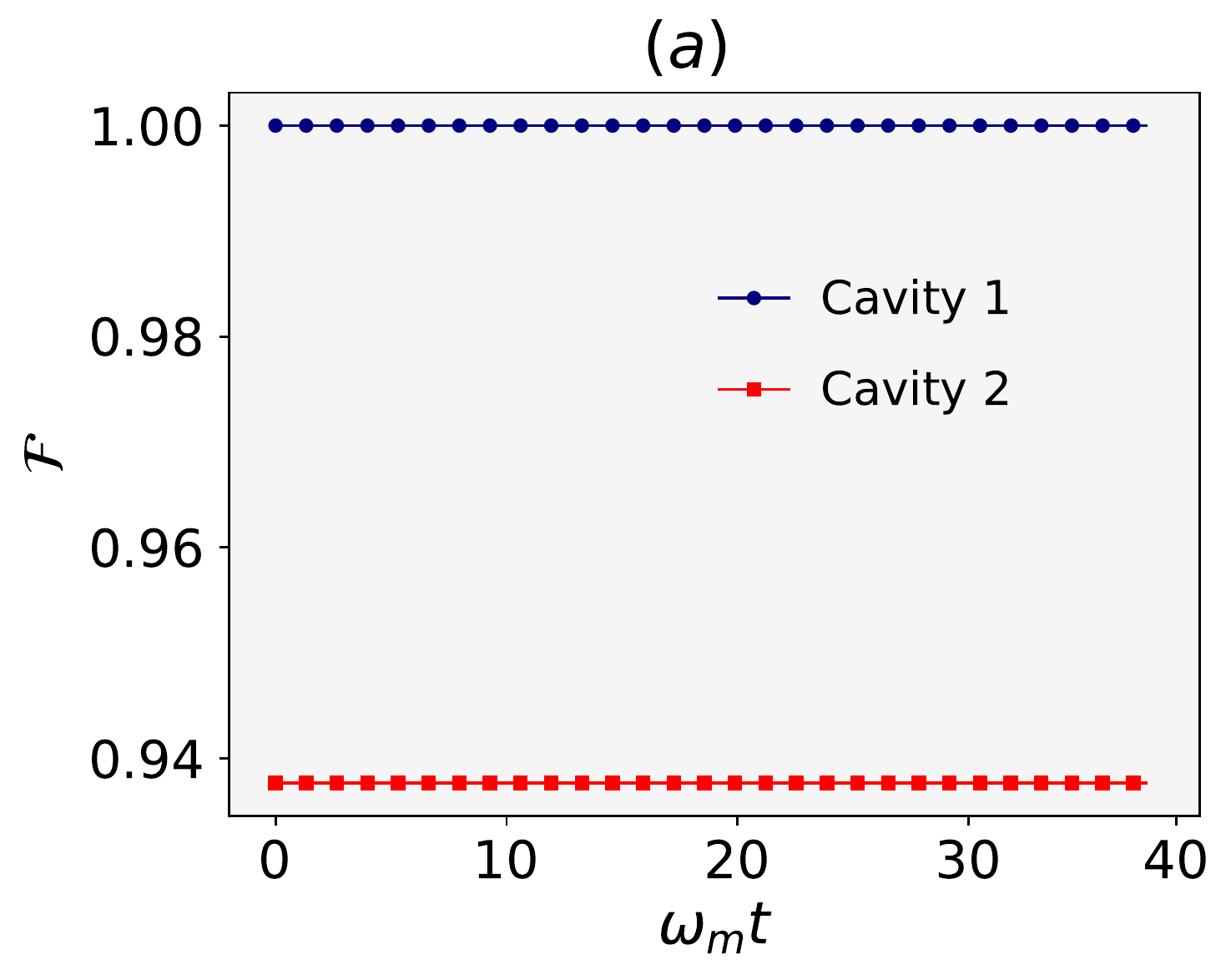}
\includegraphics[width=0.32\linewidth]{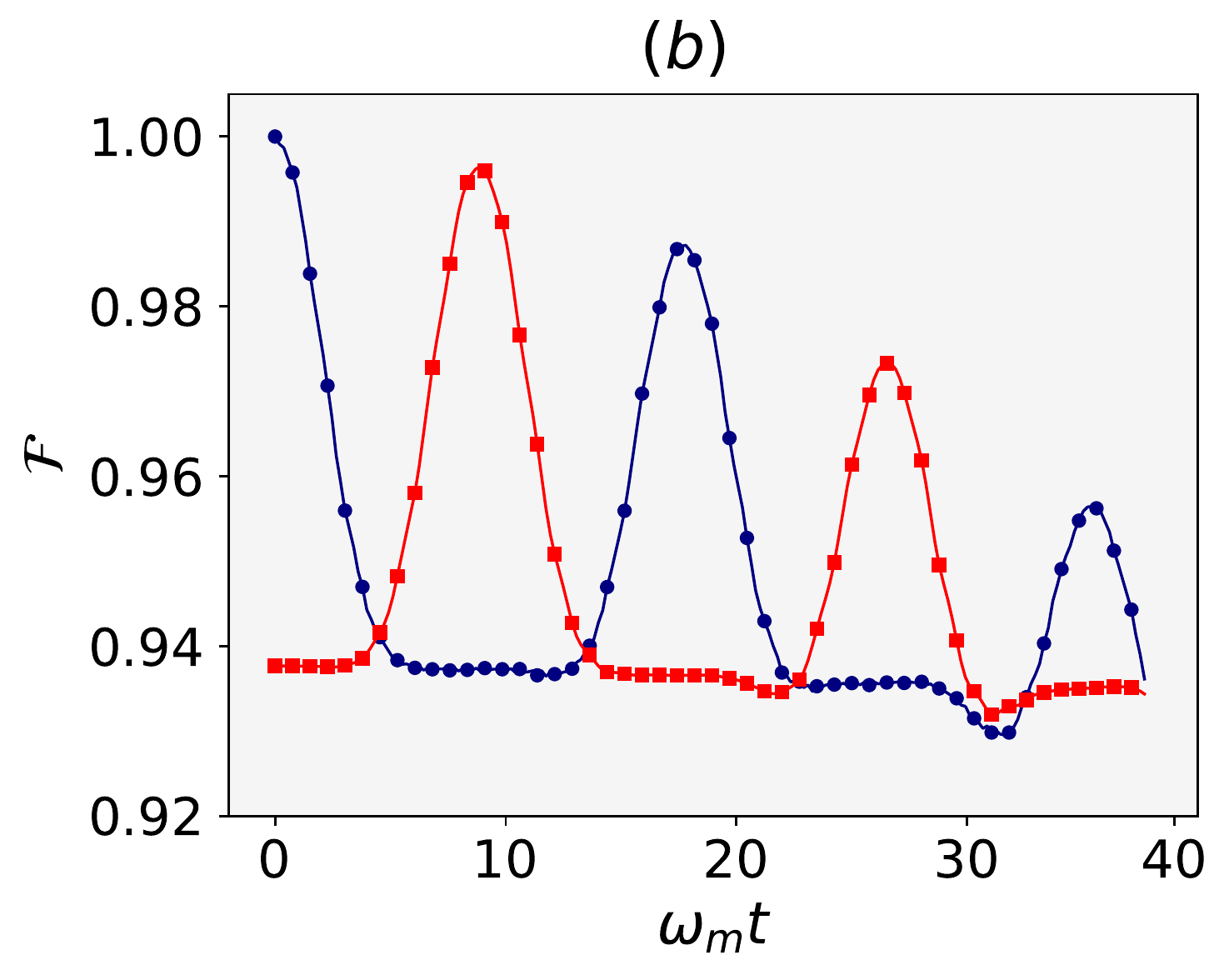}
\includegraphics[width=0.32\linewidth]{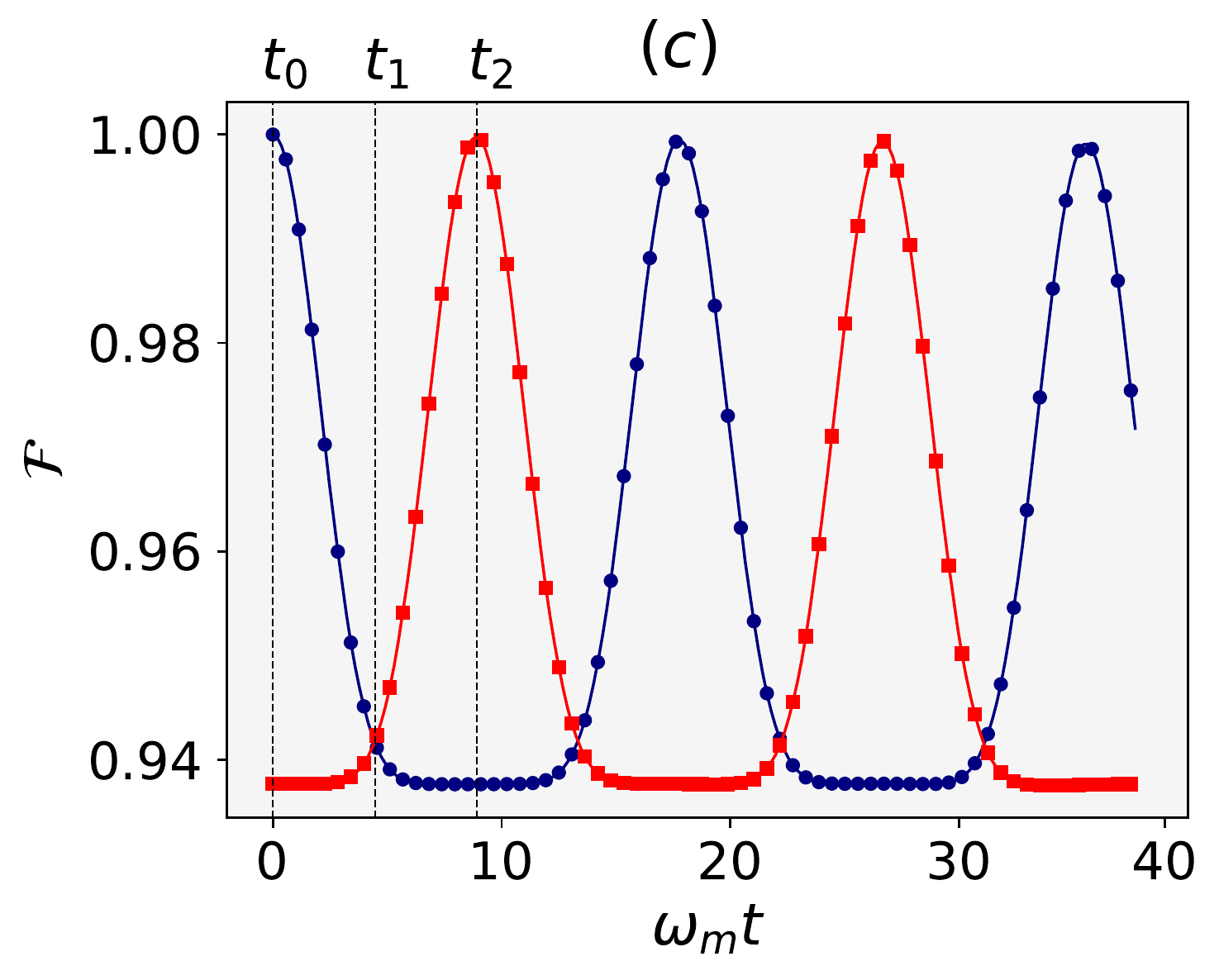}

\includegraphics[width=0.32\linewidth]{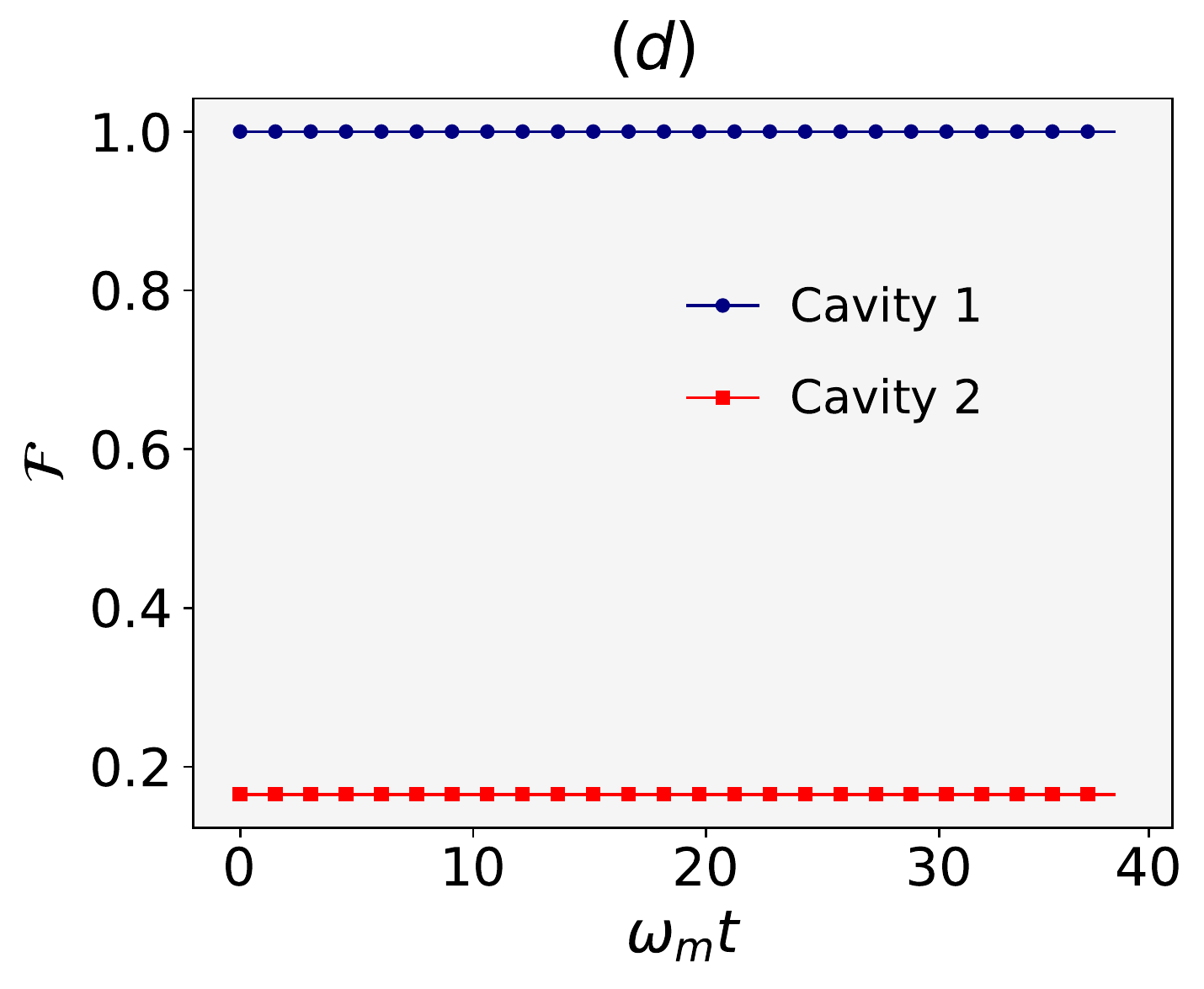}
\includegraphics[width=0.32\linewidth]{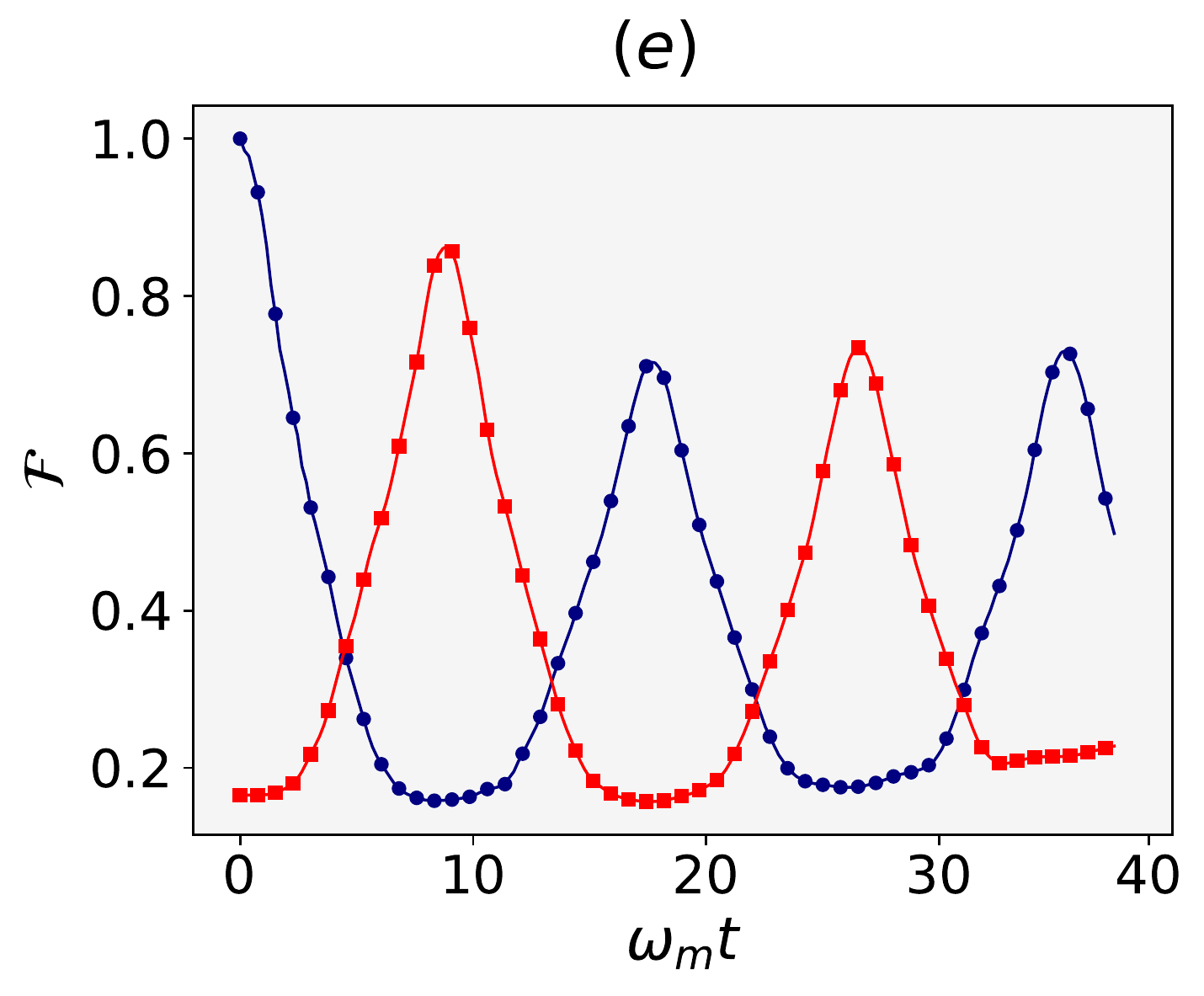}
\includegraphics[width=0.32\linewidth]{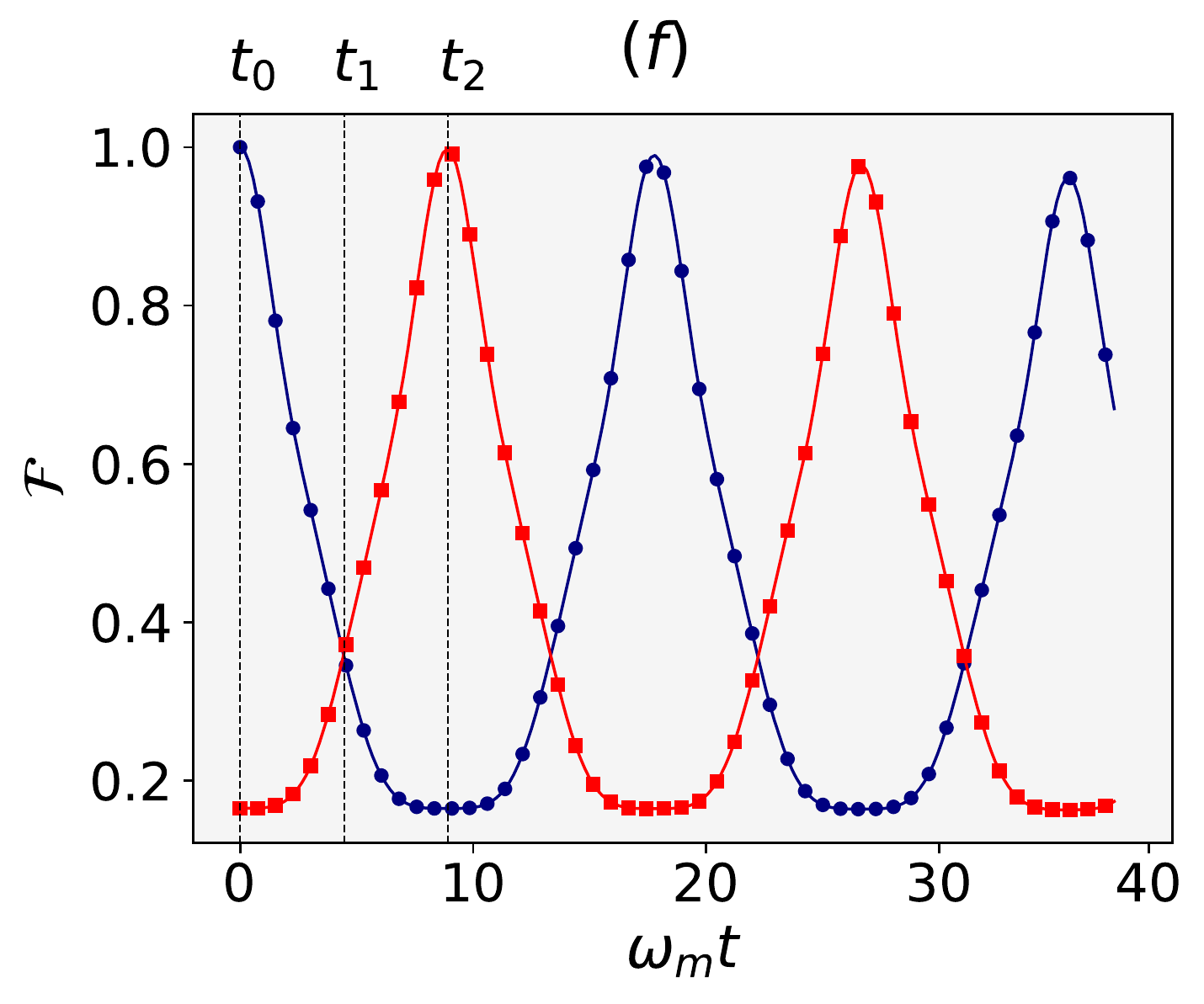}
\caption{Lossless dynamics of fidelity between the initial squeezed (top line) and cat (bottom line) state of the first cavity and other thermal bosonic modes, as second cavity and MO: $(a,d)$ $\Omega_{1}^{(j)}=\Omega_{2}^{(j)}=0$; $(b,e)$ $\Omega_{1}^{(j)}=\Omega_{2}^{(j)}=10$; $(c,f)$ $\Omega_{1}^{(j)}=\Omega_{2}^{(j)}=80$. As seen at time $t_2$ one observes a high-fidelity transfer ($\mathcal{F}\approx1$) of the squeezed state between the cavities. The parameters (in units of $\omega_{m}$) are: $g_{j}=100$, $\lambda=0.01$, $\gamma_{10}^{(j)}=\gamma_{21}^{(j)}=\kappa_{a}^{(j)}=\kappa_{b}=0$, $\xi=0.5$, $\bar{n}_{m}=\bar{n}_{c_2}=0.001$.}
\label{fig2}
\end{figure*}

\section{Dynamical transfer of quantum states between cavities}\label{sec3}

\subsection{Initialization of the cavities, mechanical oscillator and atoms}

In this section, we start with the initial conditions of the states for the two cavities, MO and the two atoms. Particularly, we will consider two different non-classical initial states for the first cavity: 

$(I)$ Squeezed state given by
\begin{equation}\label{state1}    |\psi(0)\rangle_{c_1}=\exp{\left[\frac{1}{2}\left(\xi^{*}a_{1}^{2}-\xi a_{1}^{\dagger2}\right)\right]}|0\rangle,
\end{equation}
where $\xi=r\exp{(\imath\theta)}$ is the squeezing parameter. 

$(II)$ Schr\"{o}dinger's cat state of the form
\begin{equation}\label{state2}    |\psi(0)\rangle_{c_1}=\mathcal{N}\left(|\alpha\rangle+|-\alpha\rangle\right),
\end{equation}
where $\mathcal{N}$ is a normalization constant and $|\alpha\rangle$ is a common coherent state.

Our purpose is to transfer the quantum states in Eqs. (\ref{state1}-\ref{state2}) to the second cavity, which initially is in a thermal state, similarly as the mechanical oscillator. In the coherent basis the state of the second cavity can be written as
\begin{equation}
\label{estadotermal}
    \rho_{c_2}(0)=\frac{1}{\pi\bar{n}_{c_2}}\int|\alpha\rangle\langle\alpha|e^{-\frac{|\alpha|^{2}}{\bar{n}_{c_2}}}d^{2}\alpha,
\end{equation}
and the mechanical oscillator as
\begin{equation}\label{estadomo}
    \rho_{m}(0)=\frac{1}{\pi\bar{n}_{m}}\int|\beta\rangle\langle\beta|e^{-\frac{|\beta|^{2}}{\bar{n}_{m}}}d^{2}\beta,
\end{equation}
where $\alpha$ and $\beta$ are in general complex numbers. Here $ \bar{n}_{m(c)}=\left(\exp{\left[\hbar \omega_{m(c)}/(\kappa_{B}T)\right]}-1\right)^{-1}\equiv \langle \hat{n}_{m(c)} \rangle_0$, is the average value of phonon (photons) occupation number initially in the thermal equilibrium with the reservoirs at temperature $T$, and $\kappa_{B}$ is the Boltzmann's constant. 

In addition, the two three-level atoms are initialized in the ground state
\begin{equation}
    \rho_{a_{1}}(0)=\rho_{a_{2}}(0)=|0\rangle\langle0|.
\end{equation}

\subsection{Dissipative dynamics under the Markovian master equation}
If we now include the dissipation caused by the system-environment coupling, the dissipative dynamics of the hybrid quantum system is described by the Markovian master equation (ME) for the the density matrix as follows

\begin{eqnarray}
    \label{dinamica}
    \frac{d\rho}{dt}=-\imath[\mathcal{H}_{2}+\mathcal{{H}}_{L},\rho]+\sum_{j=1}^{2}\frac{\gamma_{21}^{(j)}}{2}\left(1+\bar{n}_{a_j}\right)\mathcal{L}[\sigma_{21,j}^{-}] \nonumber \\
    + \frac{\gamma_{21}^{(j)}}{2}\bar{n}_{a_j}\mathcal{L}[\sigma_{21,j}^{+}]
    +\frac{\gamma_{10}^{(j)}}{2}\left(1+\bar{n}_{a_j}\right)\mathcal{L}[\sigma_{10,j}^{-}]\nonumber\\
    +\frac{\gamma_{10}^{(j)}}{2}\bar{n}_{a_j}\mathcal{L}[\sigma_{10,j}^{+}] + \frac{\kappa_{a}^{(j)}}{2}\left(1+\bar{n}_{c_j}\right)\mathcal{L}[a_{j}] \nonumber\\
    +\frac{\kappa_{a}^{(j)}}{2}\bar{n}_{c_j}\mathcal{L}[a_{j}^{\dagger}]+\frac{\kappa_{b}}{2}\left(1+\bar{n}_{m}\right)\mathcal{L}[b]+\frac{\kappa_{b}}{2}\bar{n}_{m}\mathcal{L}[b^{\dagger}],
\end{eqnarray}
where the common Lindblad dissipative terms are defined by:  $\mathcal{L}[\mathcal{O}]=2\mathcal{O}\rho \mathcal{O}^{\dagger}-\mathcal{O}^{\dagger}\mathcal{O}\rho-\rho \mathcal{O}^{\dagger}\mathcal{O}$. 

Here $\bar{n}_{a_j}$, $\bar{n}_{c_j}$ and $\bar{n}_{m}$ are the average occupation number for the thermal reservoirs for atoms, photons and phonons, respectively, and $\gamma_{21}^{(j)}$ $(\gamma_{10}^{(j)})$ corresponds to spontaneous emission rate from level $|2\rangle_{j}$ to $|1\rangle_{j}$  (level $|1\rangle_{j}$ to $|0\rangle_{j}$), and $\kappa_{a}^{(j)}$ $(\kappa_{b})$ is the decay rate of the $j$-cavity (mechanical) mode, respectively.

In the following sections we will numerically calculate some characteristics as fidelity, entanglement and quadrature fluctuations (QF) under the approximation of $\bar{n}_{a}=\bar{n}_{c}=\bar{n}_{m}\ll 1$. This choice is realistic considering some recent experiments, for example for a hybrid system with the MO in the regime of microwave frequencies, as in \cite{Mir2020,Rie2016} the mechanical mode may have $\omega_{m}/2\pi \approx 2$ GHz, and by cooling the system to the temperatures of $\sim 10$ mK, one gets $\bar{n}_{th}\approx 10^{-4}$.

\subsection{Fidelity of the transfer protocol}

Now, let's use the measure of fidelity as a figure of merit to quantify the efficiency of transfer of states during their evolution, which is defined as
\begin{equation}\label{fid}
    F(\rho_{c_1}(0),\rho(t))\equiv Tr\sqrt{\sqrt{\rho_{c_1}(0)}\rho(t)\sqrt{\rho_{c_1}(0)}}.
\end{equation}
where $\rho(t)\equiv\left\{\rho_{c_1}(t),\rho_{c_2}(t),\rho_{m}(t)\right\}$ define the states of first cavity, second cavity and MO, respectively. The above definition shows that the measurement is made between the initial state of the first cavity and the state of any bosonic mode during its evolution. 

In the following, we evaluate the fidelity according to definition in Eq. \ref{fid} for the lossless case, see Figs. \ref{fig2}. First, the numerical calculations indicate that, when there is no atomic pump, i.e. $\Omega^{(j)}_{1}=\Omega^{(j)}_{2}=0$, the fidelity for each bosonic mode remains constant during the time evolution, implying that the modes conserve their initial states, see Fig. \ref{fig2}$(a,d)$.
On the other hand, in  Fig. \ref{fig2}$(b,e)$ we can see how the presence of atomic pumps, e.g. $\Omega^{(j)}_{1}=\Omega^{(j)}_{2}=10$, begin to stimulate the transfer of the quantum state from the first to the second cavity. In this case, one finds that all boson modes increase their transfer probability stimulated by the atomic pump, however this pump intensity is not enough to achieve high fidelity state transfer to the second cavity. Finally, when a sufficient optimal atomic pump intensity is considered, e.g. $\Omega^{(j)}_{1}=\Omega^{(j)}_{2}=80$, it then allows at certain times the transfer at high fidelity, closed to one, see Fig. \ref{fig2}$(c,f)$. We point out that high fidelity is possible only in the first cycles of the time evolution, however one observes the fidelity will decreases in time because the three modes becomes more and more entangled and this effect destroy the periodic transfer, see Fig. \ref{fig5} and the corresponding analysis in the next section. 
\begin{figure}[t]
\centering
\includegraphics[width=0.5\linewidth]{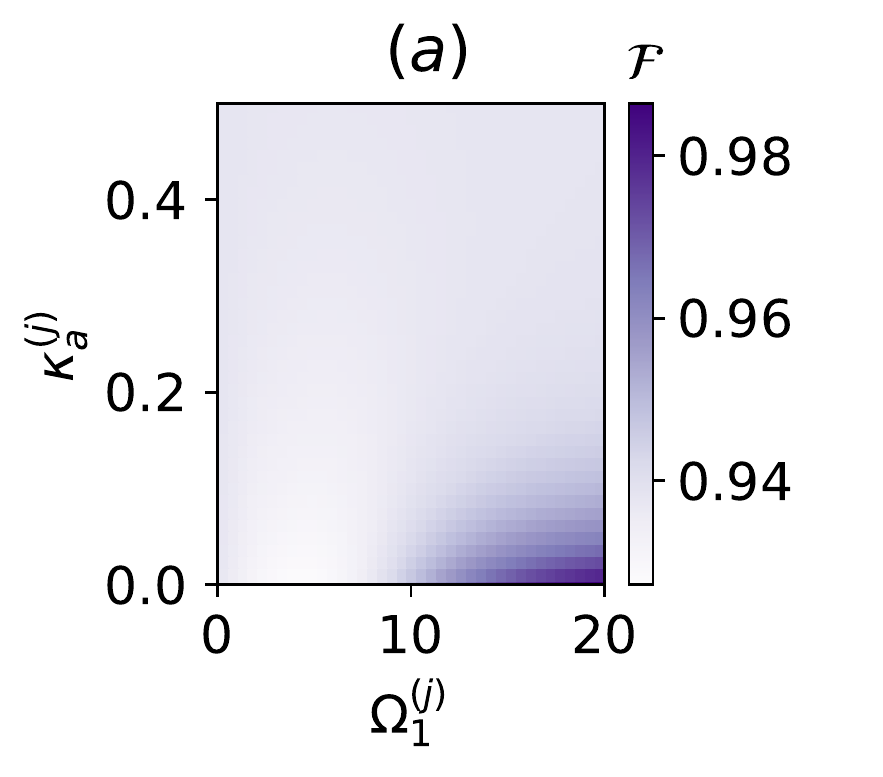}
\includegraphics[width=0.48\linewidth]{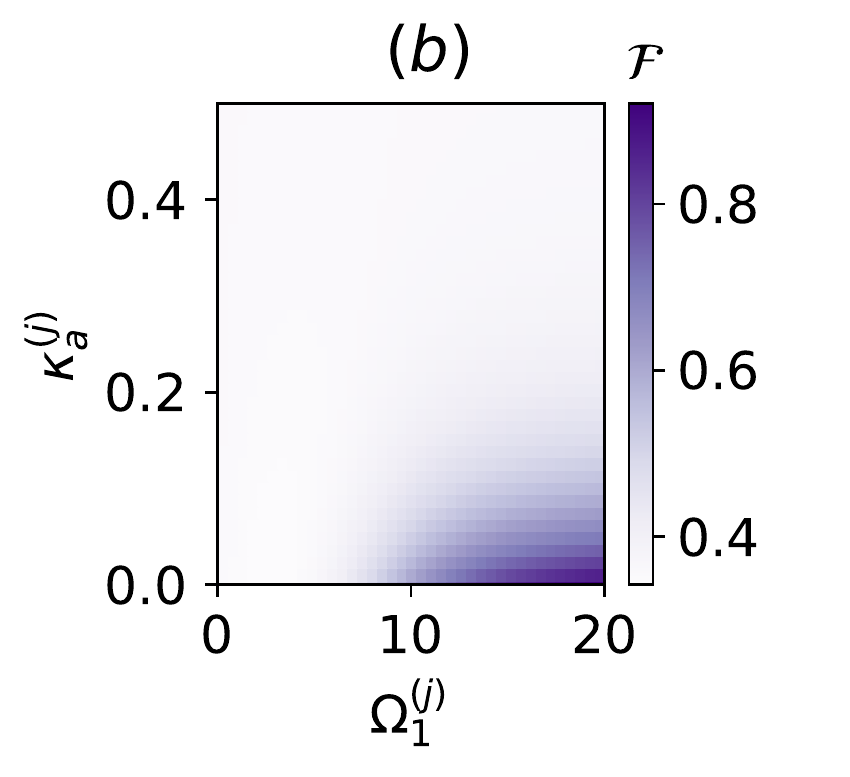}
\caption{Fidelity of transferring: $(a)$ squeezed state ($\xi=0.5$) and $(b)$ cat state ($\alpha=2$), from the first cavity to the second cavity, calculated at the instant $t_{2}$ (see Fig.\ref{fig2}) as a functions of cavity losses $\kappa^{(j)}_{a}$ and the atomic pump $\Omega^{(j)}_{1}$. Here, we have considered $\Omega^{(j)}_{2}=\Omega^{(j)}_{1}$ and $\kappa^{(j)}_{b}=0.01\kappa^{(j)}_{a}$. Other parameters (in units of $\omega_{m}$) are: $g_{j}=100$, $\lambda=0.01$, $\gamma_{21}^{(j)}=0$, $\gamma_{10}^{(j)}=20$, and $\bar{n}_{m}=\bar{n}_{a_j}=\bar{n}_{c_j}=10^{-3}$.}
\label{fig3}
\end{figure}
Therefore, when the atomic system is under the external pump, the transfer of the quantum state between cavities becomes possible with a fidelity closed to one. The results in Fig. \ref{fig2} correspond to an ideal situation, i.e. for unitary dynamics, but the realistic systems are exposed to dissipation and decoherence, therefore, we carried out a study taking into account the influence of the dissipation rates of atoms, cavities and MO on the fidelity of the transfer of state between the cavities. On the other hand, the effect of the dissipation one could compensate by the driving fields. In Fig. \ref{fig3}, we evaluate the fidelity between the first and second cavity field at the dimensionless time $t_{2}$ (see Fig. \ref{fig2}c) as a function of the atomic driving strength such that $\Omega^{(j)}_{1}=\Omega^{(j)}_{2}$, and the cavity damping rates, $\kappa^{(j)}_{a}$, considering a fixed dissipation rate for the atoms, e.g. $\gamma^{(j)}_{10}=20 \omega_m$, and the damping rate for the MO is $\kappa_{b}=0.01 \kappa_{a}$. We observe that the optimal fidelity exists when the driving increases and the losses decrease, as it is achievable by the available means.

\section{Quantum entanglement} \label{sec4}
 In this section, we study the bipartite and tripartite quantum entanglement for our hybrid system. Moreover, we analyze the relation between the entanglement and the transfer effect, studied in the previous section.

In general, for subsystems $A$ and $B$ and an associated density matrix $\hat{\rho}_{AB}$, the negativity \cite{VIDAL2002} is defined as 
\begin{equation}
    \mathcal{N}(\hat{\rho}_{AB})=\sum_{i}\frac{|\zeta_{i}|-\zeta_{i}}{2},
\end{equation}
where $\zeta_{i}$ are the eigenvalues of the partial transpose of the density matrix $\hat{\rho}_{AB}$ with respect to one of the subsystems.

In addition, we employ the measure of genuine tripartite entanglement, as the minimum residual contangle \cite{ADESSO2006}, defined as

\begin{equation}
    E_{l}^{A|B|C}=\min_{A,B,C}\left[E_{l}^{A|(BC)}-E_{l}^{A|B}-E_{l}^{A|C}\right]
\end{equation}

where the contangles $\left\{E_{l}^{A|(BC)}, E_{l}^{A|B}, E_{l}^{A|C}\right\}$ are defined as the quadratic logarithm of $\left\{\|\hat{\rho}^{T_{A}}\|, \|\hat{\rho}^{T_{A}}_{AB}\|, \|\hat{\rho}^{T_{A}}_{AC}\|\right\}$ with the trace norm $(\|\cdot\|)$, partial transpose (superscript), and partial trace (subscript).

\begin{figure}[t]
\centering
\includegraphics[width=0.83\linewidth]{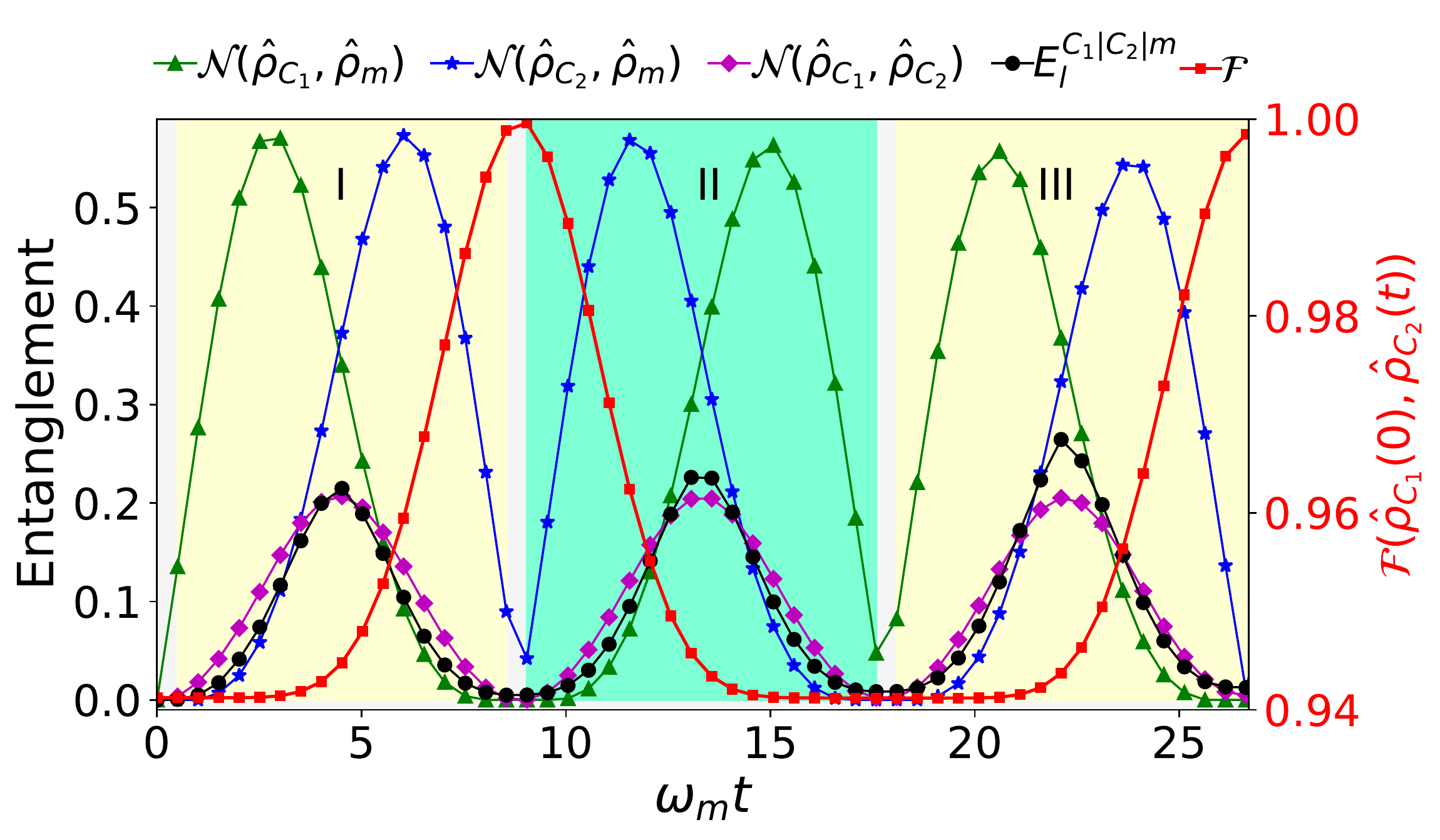}
\caption{Lossless time evolution of quantum entanglement (left-hand black axis) between the bosonic modes when the squeezing is induced in the first cavity and Fidelity (right-hand red axis) between the initial state of cavity 1 and evolved state of the cavity 2. Other parameters (in units of $\omega_{m}$) are: $\lambda=0.01$, $g_{j}=100$, $\bar{n}_{a,j}=\bar{n}_{c,j}=\bar{n}_{m}=0.001$, $\Omega_{1}^{(j)}=\Omega_{2}^{(j)}=80$, $\gamma_{10}^{(j)}=\gamma_{21}^{(j)}=\kappa_{a}^{(j)}=\kappa_{b}=0$.}
\label{fig4}
\end{figure}

\begin{figure*}[t]
\centering
\includegraphics[width=0.32\linewidth]{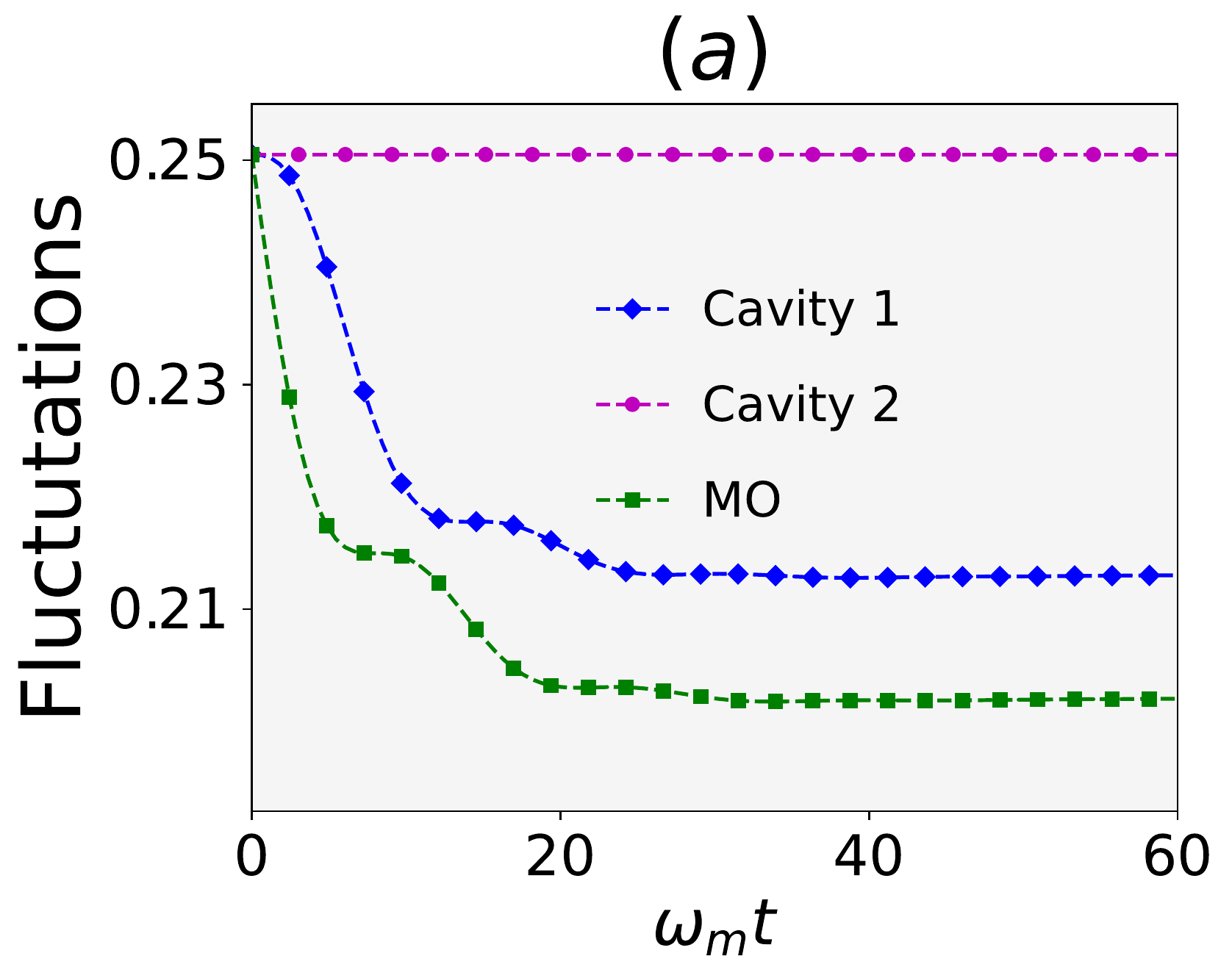}
\includegraphics[width=0.32\linewidth]{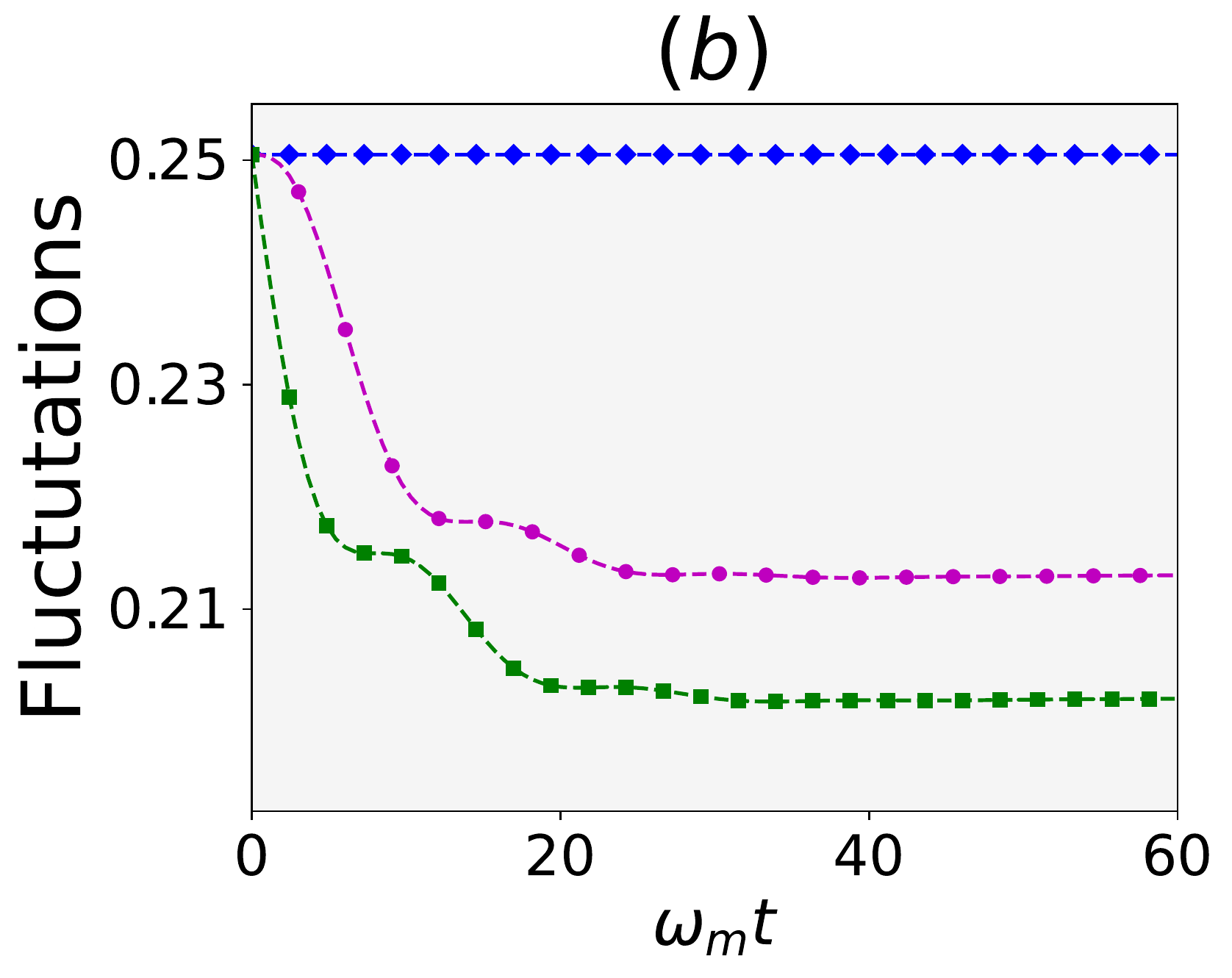}
\includegraphics[width=0.32\linewidth]{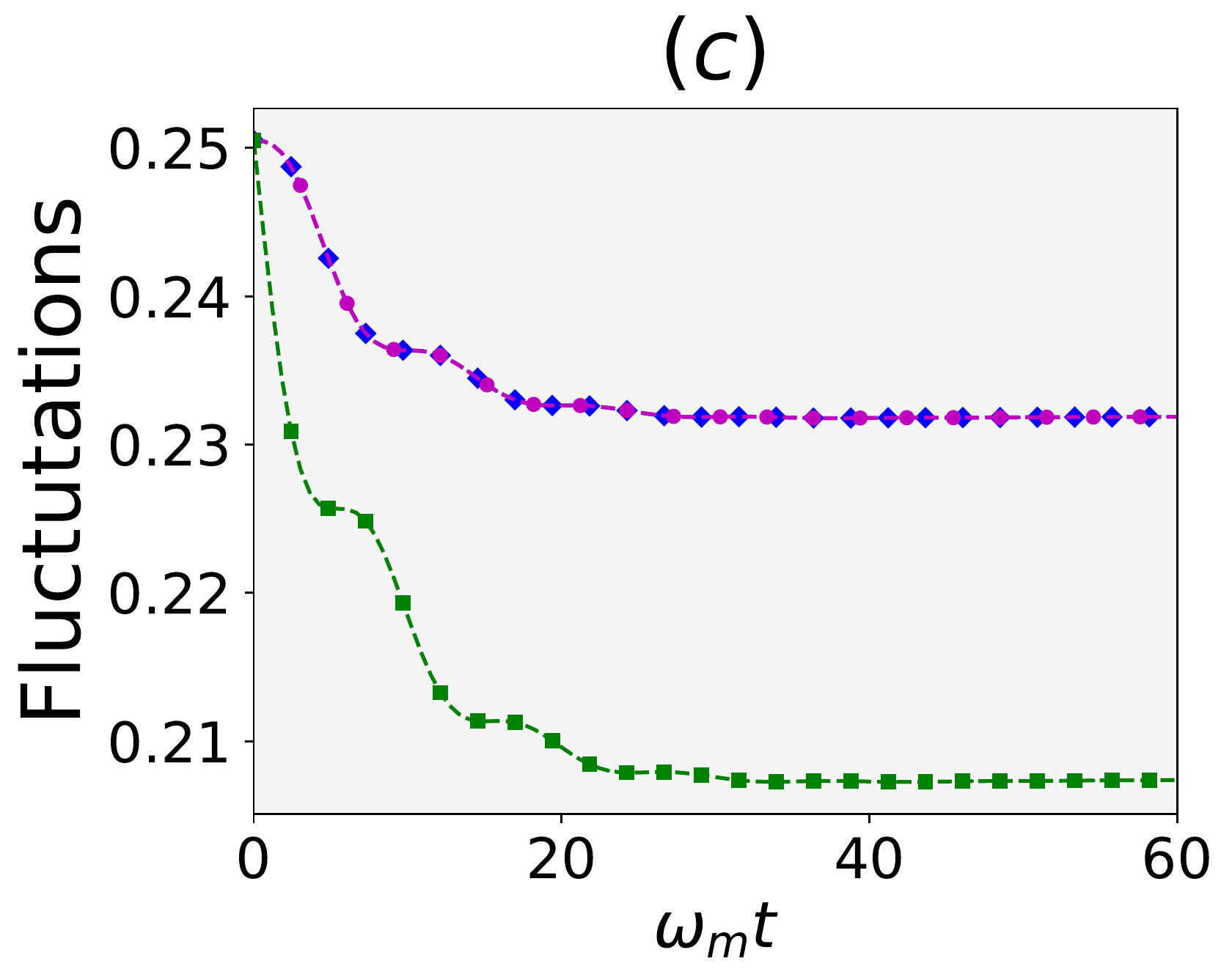}

\includegraphics[width=0.32\linewidth]{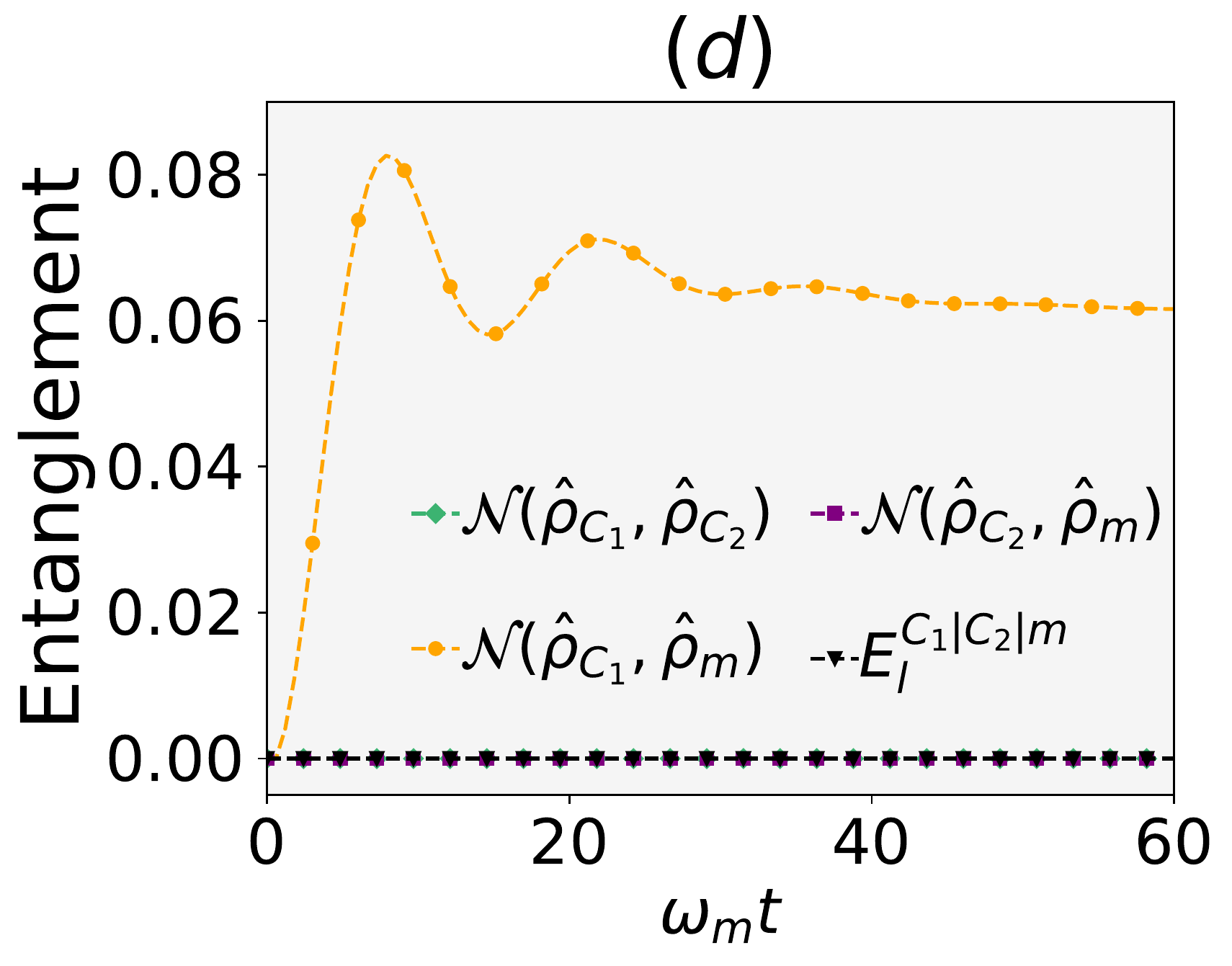}
\includegraphics[width=0.32\linewidth]{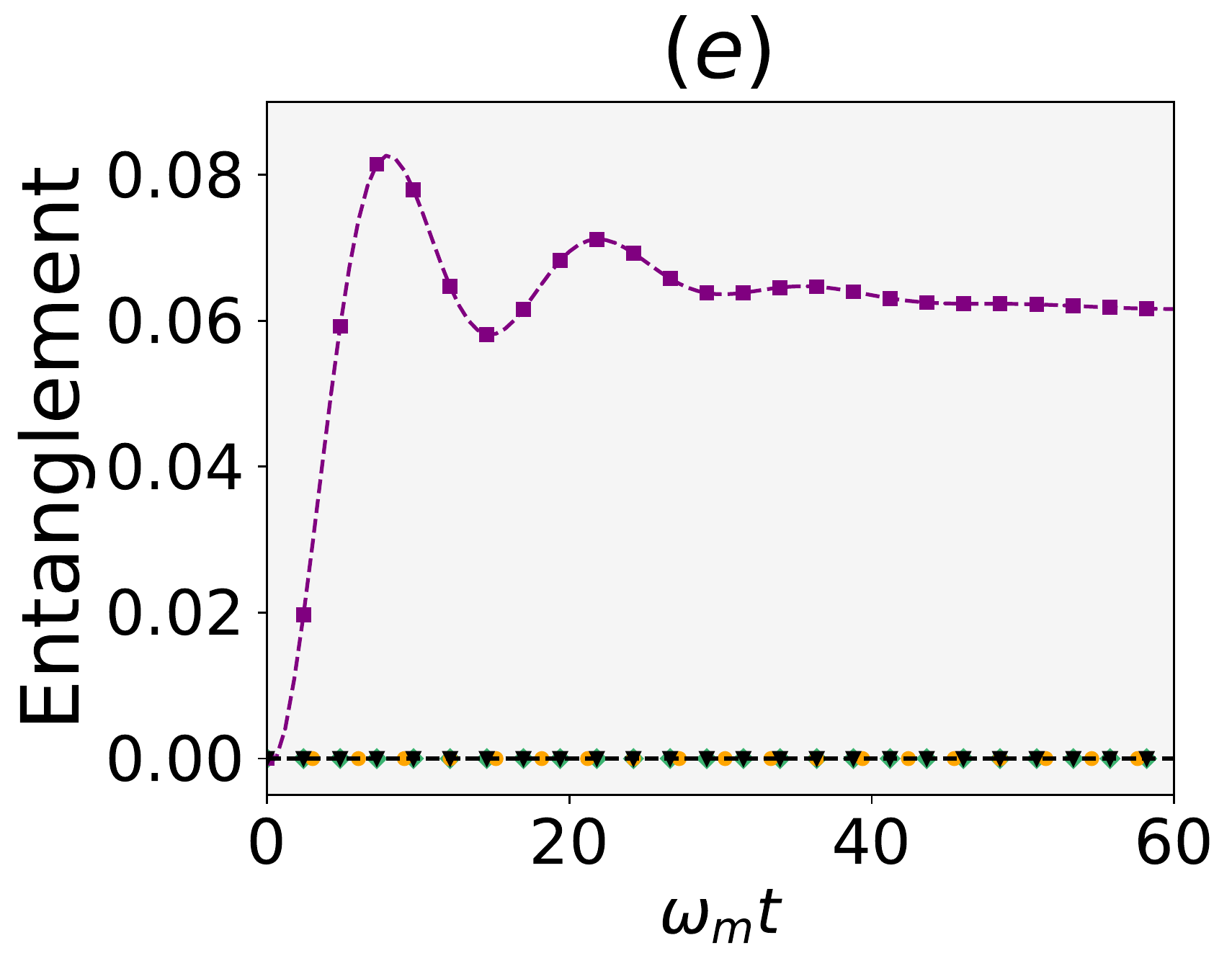}
\includegraphics[width=0.32\linewidth]{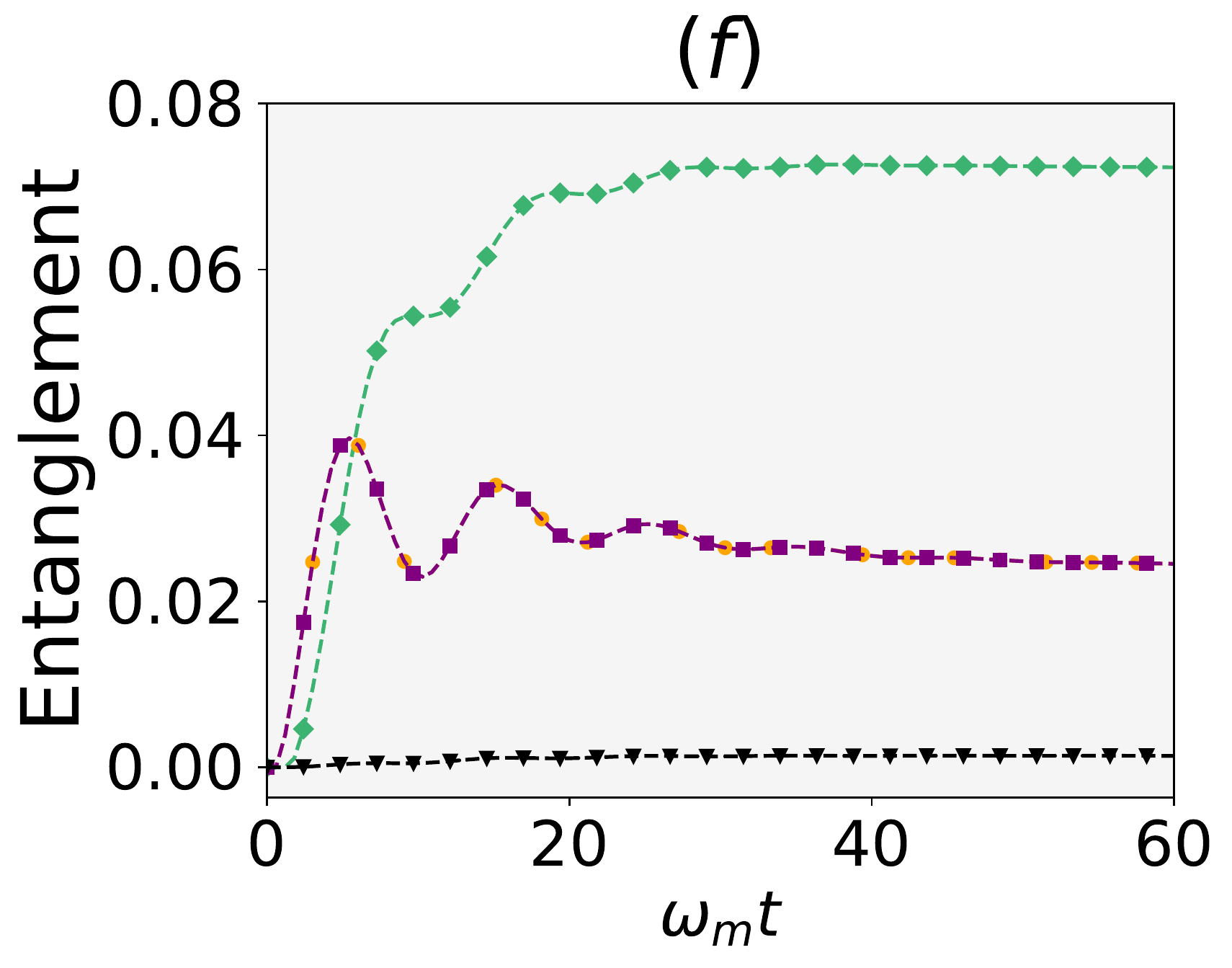}
\caption{Dissipative time evolution of Quantum Fluctuations (top panel) and Entanglement (bottom panel) of the bosonic modes when the squeezing is induced in the MO. Managing the Jaynes-Cummings couplings so that the tripartite couplings are: $(a,d)$ $\Lambda_{1}=1$, $\Lambda_{2}=0$, $(b,e)$ $\Lambda_{1}=0$, $\Lambda_{2}=1$, $(c,f)$ $\Lambda_{1}=\Lambda_{2}=1$. Other parameters are: $\bar{n}_{a,j}=\bar{n}_{c,j}=\bar{n}_{m}=0.001$, $q=0.01$, $\Omega_{1}^{(j)}=\Omega_{2}^{(j)}=20$, $\gamma_{10}^{(j)}=20$, $\gamma_{21}^{(j)}=0$, $\kappa_{a}^{(j)}=0.2$, $\kappa_{b}=0.01\kappa_{a}$, $\phi_{c_{1}}=\phi_{c_{2}}=\pi/4$, $\phi_{m}=-\pi/4$.}
\label{fig5}
\end{figure*}

In stage I of Fig. \ref{fig5} we can see that initially bipartite entanglement is generated between cavity 1 and MO (see green curve), and later in time MO will be entangled with the second cavity (see blue curve). On the other hand, a minor bipartite entanglement is generated between the cavities (see magenta curve). Its maximum value occurs when the other bipartite entanglements are equal. Additionally, the tripartite entanglement is generated, which becomes maximum for this same time (see black curve).
In this first stage we can conclude that when the bipartite entanglements $\{\mathcal{N}(\rho_{C_1, MO}), \mathcal{N}(\rho_{C_2, MO})\}$ vanish, the transfer of states between the cavities occurs with the maximal fidelity (see red curve). As the entanglement between the MO and cavity 2 is generated, so the initial state of cavity 1 can be transmitted to the second cavity.

In stage II we can observe that the entanglement between cavity 2 and the MO increases again, this fact produces a reduction in the transfer fidelity that finally falls when the entanglement between cavity 1 and the MO reaches a new maximum value. Here, the tripartite entanglement takes a higher maximum value than in stage I. This maximal value will increase in time as the three subsystems become more and more entangled. Therefore, one obtains additional quantum resource in our model related to the generation of tripartite entanglement which increases while the transfer efficiency and the bipartite correlations decrease over the time.

\begin{figure*}[t]
\centering
\includegraphics[width=0.295\linewidth]{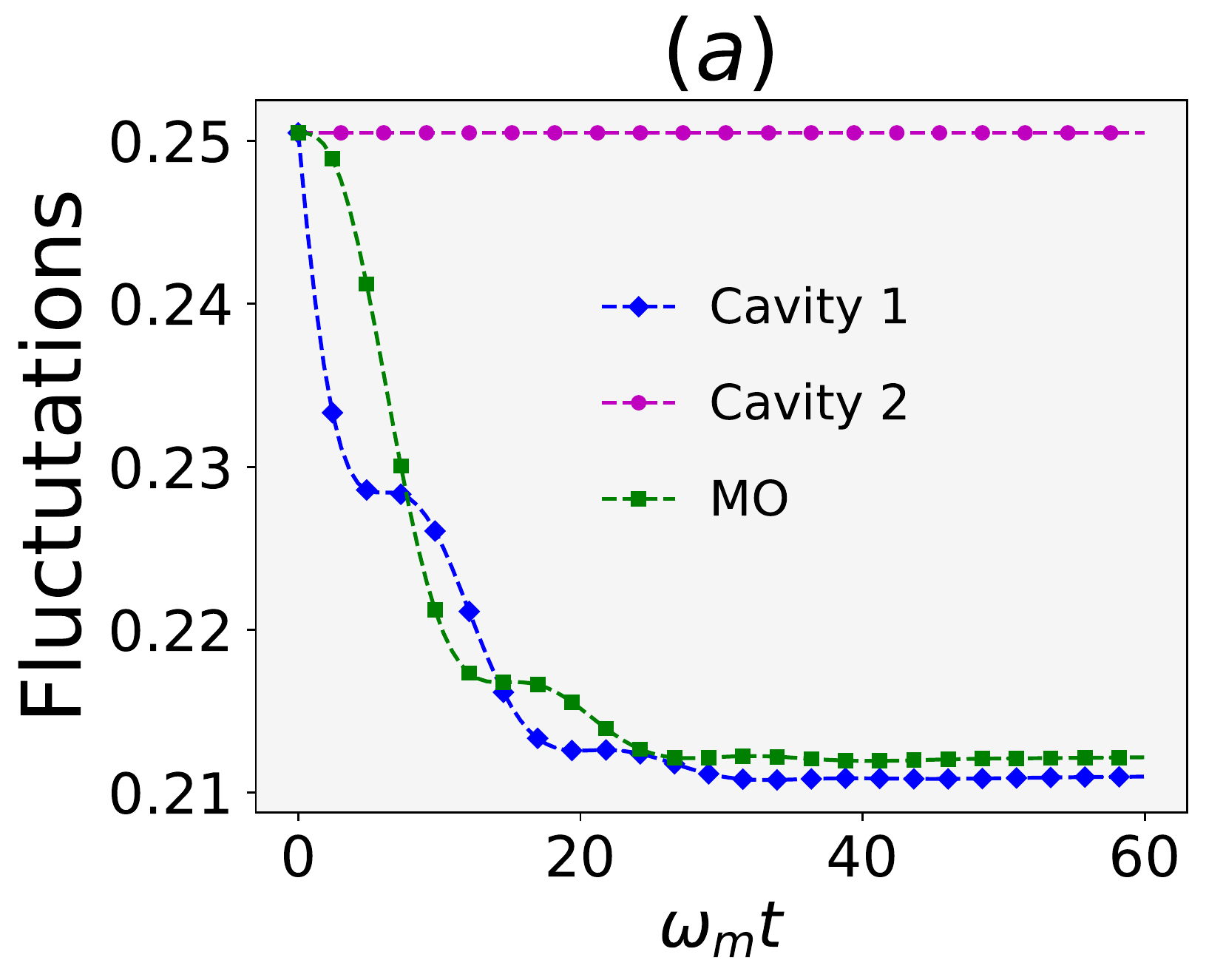}
\includegraphics[width=0.305\linewidth]{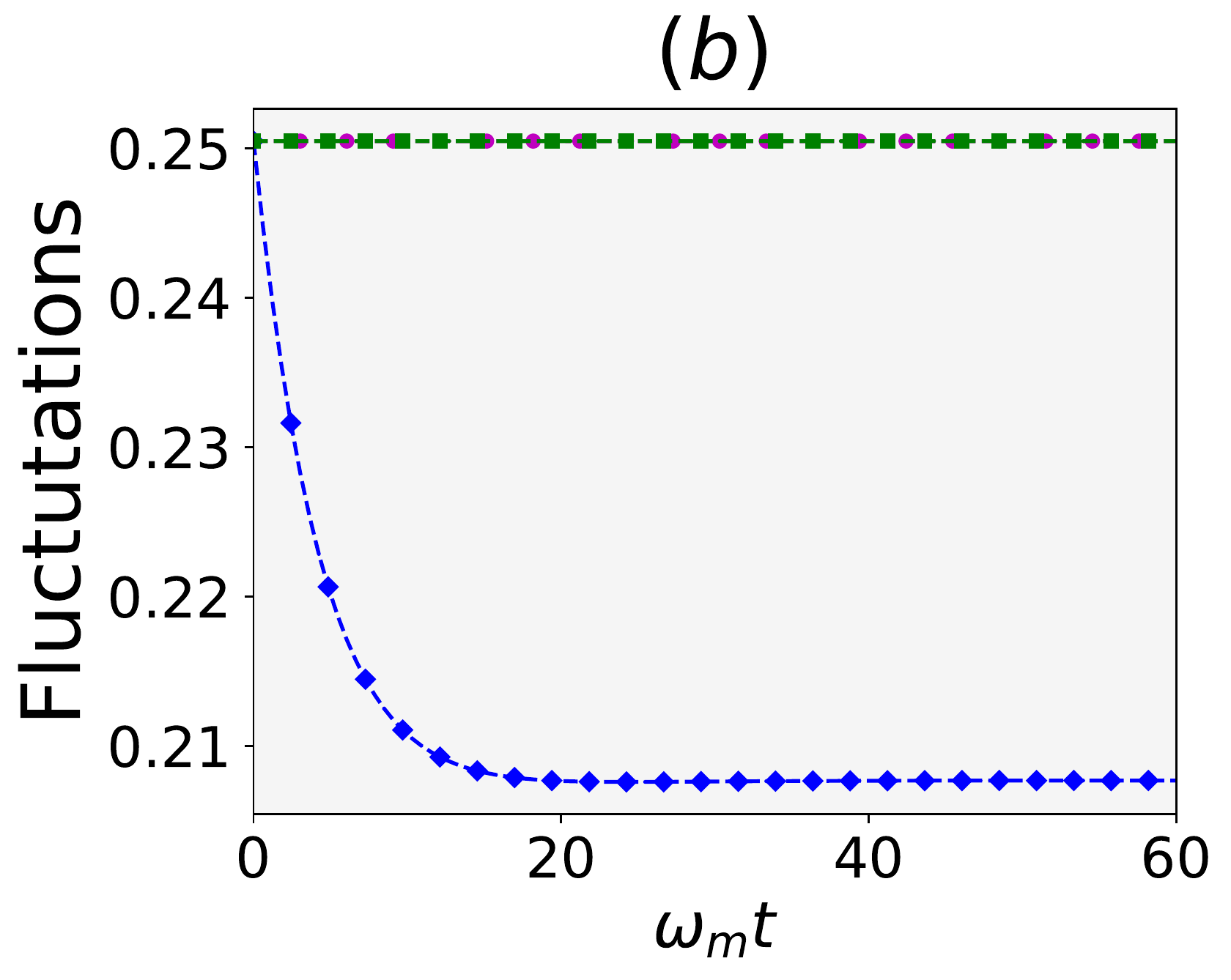}
\includegraphics[width=0.315\linewidth]{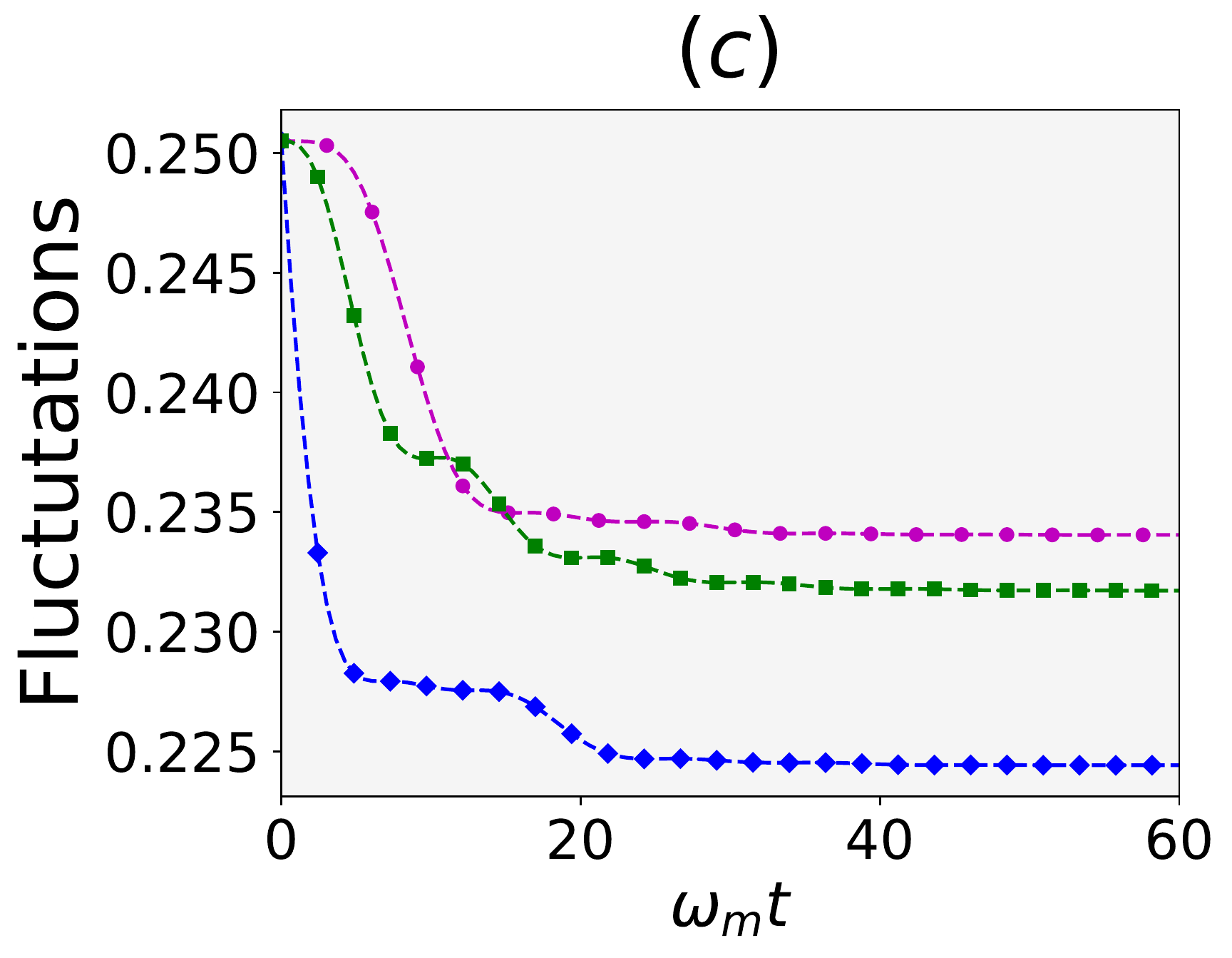}

\includegraphics[width=0.3\linewidth]{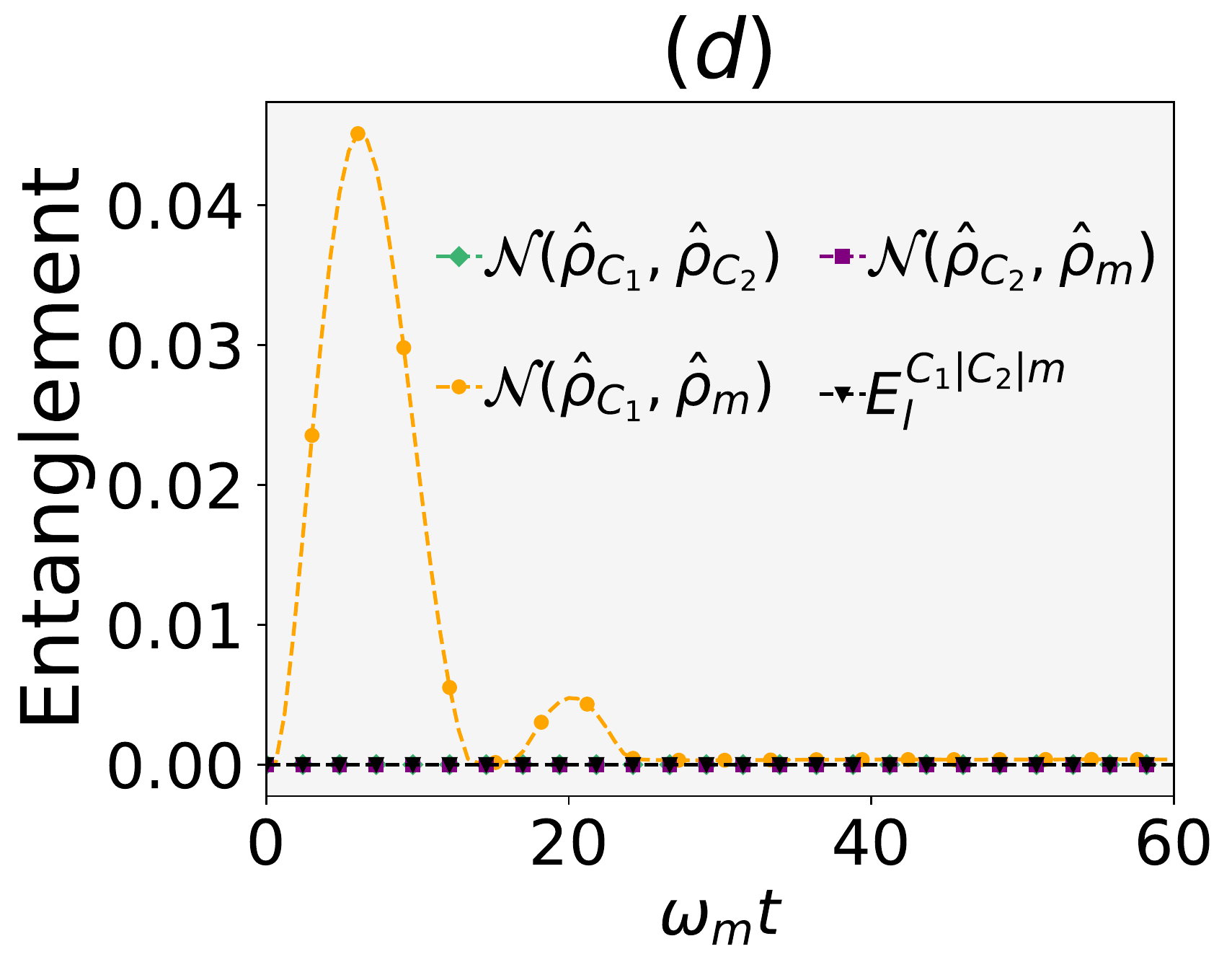}
\includegraphics[width=0.29\linewidth]{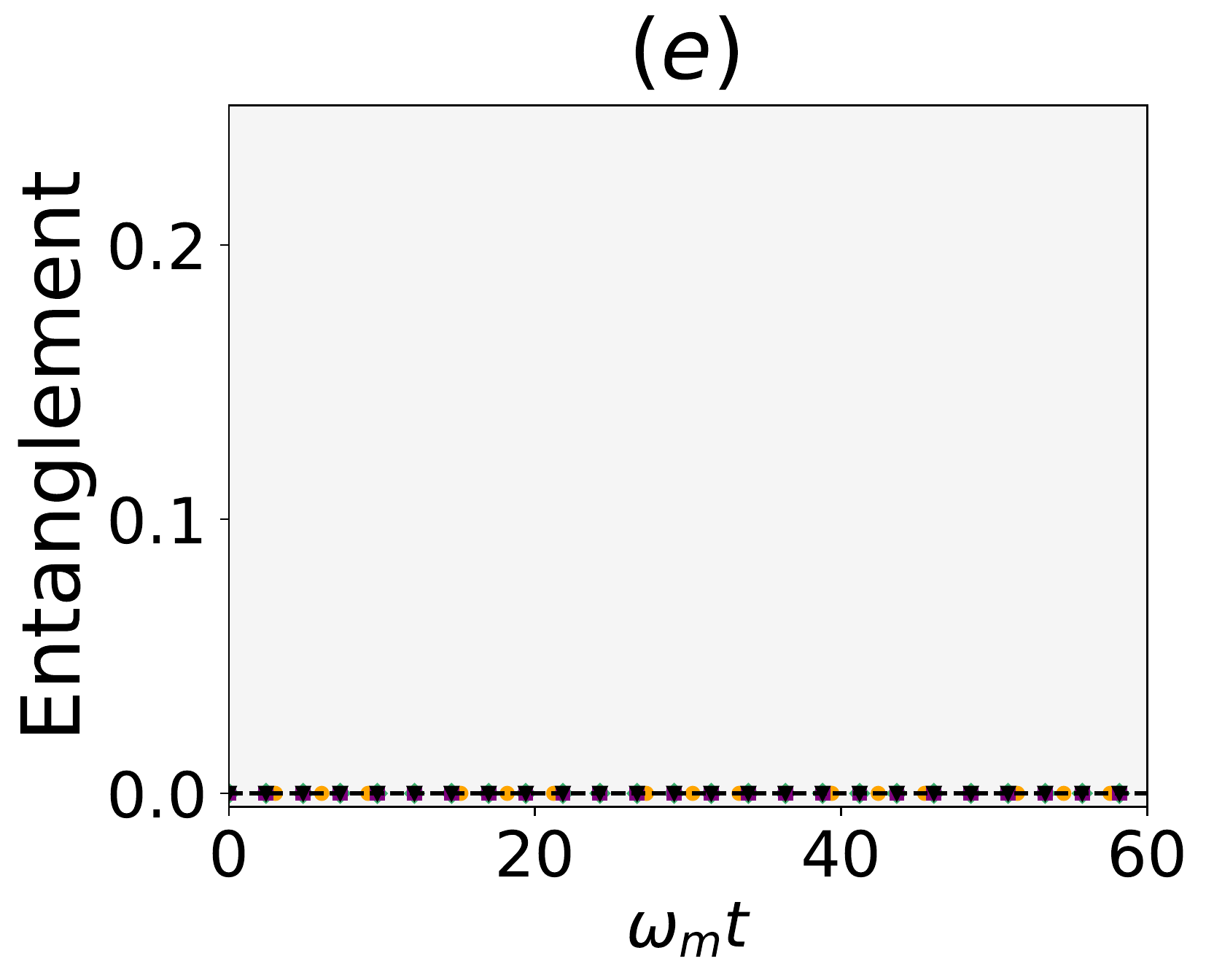}
\includegraphics[width=0.305\linewidth]{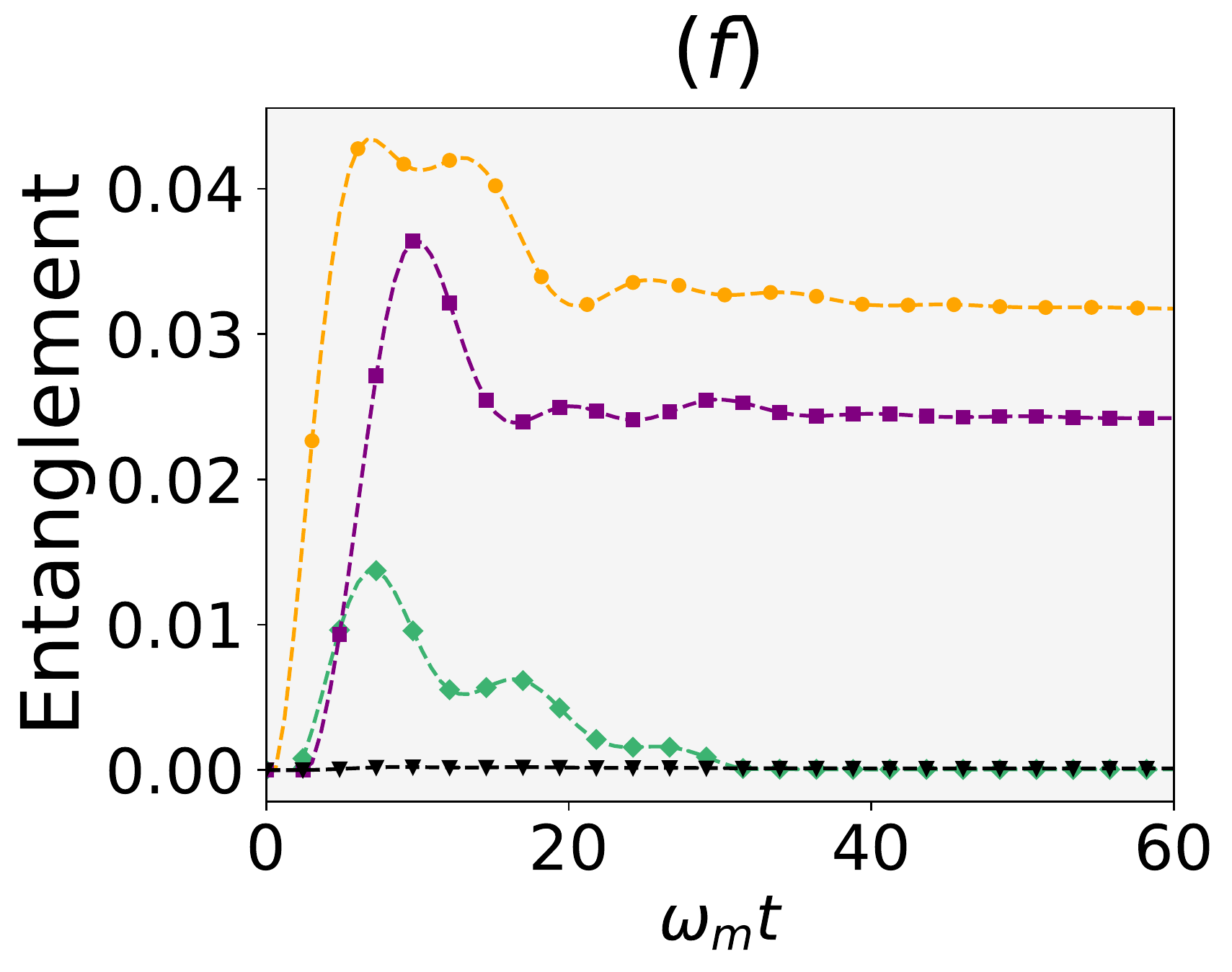}
\caption{Dissipative time evolution of QF (top panel) and entanglement (bottom panel) of the bosonic modes when the squeezing is induced in the first cavity. Managing the Jaynes-Cummings couplings so that the tripartite couplings are: $(a,d)$ $\Lambda_{1}=1$, $\Lambda_{2}=0$; $(b,e)$ $\Lambda_{1}=0$, $\Lambda_{2}=1$; $(c,f)$ $\Lambda_{1}=1, \Lambda_{2}=1.1$. Other parameters are the same as in Fig.\ref{fig5}. Additionally with $\phi_{c_{1}}=\phi_{c_{2}}=\pi/4$, $\phi_{m}=-\pi/4$.}
\label{fig6}
\end{figure*}

Finally, at the beginning of the stage III the cavity 1 has returned to its initial state, and then the cycle of the stage I is repeated. 

Concerning the case of Schr\"{o}dinger's cat transfer, we have considered an even cat state (Eq. \ref{state2}), thus containing only even Fock state terms:
\begin{equation}\label{state3}    |\psi(0)\rangle\propto2\exp{\left\{-|\alpha|^{2}/2\right\}}\sum_{n=0}^{\infty}\frac{\alpha^{2n}}{\sqrt{2n!}}|2n\rangle.
\end{equation}
Since the shape of the above state is proportional to pairs of excitations, similar to a squeezed state, we can conclude that the evolution of the quantum correlations as bipartite and tripartite entanglement will result analogous to the effect shown in Fig. \ref{fig4}. Due to this, we consider irrelevant to present here these numerical results. 

\section{Stationary synchronization of squeezing in the hybrid network}\label{sec5}
The effect studied in the previous section occurs periodically, i.e. there are definite moments when the transfer occurs. Therefore, the transfer of the quantum state is a reversible effect. On the other hand, in this section, we study the possibility of the steady-state squeezing synchronization in several modes, which we call 'quantum state synchronization' effect for the cavities and MO modes. This effect is irreversible as compared to the transfer, principally because in this case the tripartite system reaches an equilibrium between the pump and losses mechanisms, similar to the laser/maser model.

In order to realize a steady-state squeezing in various bosonic modes (cavities and MO) one should ignite the squeezing of any of these modes. Therefore, as it was observed in our recent study \cite{PRA2022}, one can stimulate initially squeezing by connecting the hybrid system to a squeezed bath. On the other hand, it is also possible to induce the squeezing by a coherent driving of a photonic or phononic mode similar to the proposal in \cite{Walls}. In the interaction picture, for example, a phonon squeezing pump is described by the Hamiltonian

\begin{equation}
    \mathcal{H}_{q}=q (b^{\dagger2}+b^{2}),
\end{equation}
where $q$ is proportional to the driving field strength. As result, the mechanical resonator can be prepared dynamically in a squeezed state. 

If we consider the system-environment interaction, the dissipative dynamics of the hybrid quantum system is described by the Markovian master equation in a Lindblad form
\begin{eqnarray}
\label{dinamica2}
    \frac{d\rho}{dt}&=&-\imath[\mathcal{H}_{2}+\mathcal{H}_{q}+\mathcal{H}_{L},\rho]+\hat{L}(\rho),
\end{eqnarray}
where $\hat{L}(\rho)$ is the losses part as appear in Eq. \ref{dinamica}.

\subsection{Definition of Quantum Fluctuations}
In the following we will calculate the degree of squeezing present in the states of the cavity and mechanical oscillator. For this, we rely on numerical methods according to \cite{qutip} to solve Eq. \ref{dinamica2} in the steady-state, i.e. $\dot{\rho}=0$, and calculate the quantum fluctuations (QF) defined by
\begin{align}\label{cuadrat222}
    \langle\left(\Delta \mathcal{X}_{\mathcal{O}}\right)^{2}\rangle=\langle\mathcal{X}^{2}\rangle-\langle\mathcal{X}\rangle^{2},
    \langle\left(\Delta \mathcal{Y}_{\mathcal{O}}\right)^{2}\rangle=\langle\mathcal{Y}^{2}\rangle-\langle\mathcal{Y}\rangle^{2},
\end{align}
with the quadratures $\mathcal{X} =\left(\mathcal{O}e^{\imath\phi_{\mathcal{O}}}+\mathcal{O}^{\dagger}e^{\imath\phi_{\mathcal{O}}}\right)/2$ and $\mathcal{Y} =\left(\mathcal{O}e^{-\imath\phi_{\mathcal{O}}}-\mathcal{O}^{\dagger}e^{\imath\phi_{\mathcal{O}}}\right)/2\imath$. Here, $\mathcal{O}$ can be a photon or phonon operator and  $\phi_{\mathcal{O}}$ permits to generalize the direction of the QF, i.e. indicating the squeezing along any pair of axes $(x', y')$ in the phase space. Then, the squeezing condition for the quadrature e.g. $\mathcal{X}$ corresponds to the relation $\langle\left(\Delta \mathcal{X}_{\mathcal{O}}\right)^{2}\rangle<0.25$.

\subsection{Influence of the squeezing pump strength and the tripartite hybrid interaction}

In this section, we study the influence of the squeezing driving of the first cavity and MO modes on the squeezing synchronization effect and how this result is correlated to the bipartite and tripartite quantum entanglement between the bosonic modes. Also the importance of the tripartite coupling on these resources is studied.
\begin{table*}[ht]
\begin{tabular}{c|cc|ccc|ccc|ccc|}
\cline{2-12}
\multicolumn{1}{l|}{}      & \multicolumn{2}{c|}{Strength}                            & \multicolumn{3}{c|}{Phase QF ($\pi/4$)}                                                                               & \multicolumn{3}{c|}{Synchronization}                                            & \multicolumn{3}{c|}{Entanglement}                                            \\ \hline
\multicolumn{1}{|c|}{Pump} & \multicolumn{1}{c|}{$\hspace{0.1cm}\Lambda_{1}\hspace{0.1cm}$} & $\Lambda_{2}$ & \multicolumn{1}{c|}{$\hspace{0.1cm}\phi_{c_{1}}\hspace{0.1cm}$} & \multicolumn{1}{c|}{$\hspace{0.1cm}\phi_{c_{2}}\hspace{0.1cm}$} & $\hspace{0.1cm}\phi_{m}\hspace{0.1cm}$ & \multicolumn{1}{c|}{$C_{1}$/$C_{2}$} & \multicolumn{1}{c|}{$C_{1}$/MO}  & $C_{2}$/MO  & \multicolumn{1}{c|}{$C_{1}$/$C_{2}$} & \multicolumn{1}{c|}{$C_{1}$/MO}  & $C_{2}$/MO  \\ \hline
\multicolumn{1}{|c|}{}     & \multicolumn{1}{c|}{1}                & 0                & \multicolumn{1}{c|}{$+$}                             & \multicolumn{1}{c|}{$+$}        & $-$    & \multicolumn{1}{c|}{}                  & \multicolumn{1}{c|}{$\checkmark$} &              & \multicolumn{1}{c|}{}                  & \multicolumn{1}{c|}{$\checkmark$} &              \\ \cline{2-12} 
\multicolumn{1}{|c|}{MO}    & \multicolumn{1}{c|}{0}                & 1                & \multicolumn{1}{c|}{$+$}                             & \multicolumn{1}{c|}{$+$}        & $-$   & \multicolumn{1}{c|}{}                  & \multicolumn{1}{c|}{}             & $\checkmark$ & \multicolumn{1}{c|}{}                  & \multicolumn{1}{c|}{}             & $\checkmark$ \\ \cline{2-12} 
\multicolumn{1}{|c|}{}     & \multicolumn{1}{c|}{1}                & 1                & \multicolumn{1}{c|}{$+$}                             & \multicolumn{1}{c|}{$+$}        & $-$   & \multicolumn{1}{c|}{$\checkmark$}      & \multicolumn{1}{c|}{}             &              & \multicolumn{1}{c|}{$\checkmark$}      & \multicolumn{1}{c|}{$\checkmark$} & $\checkmark$ \\ \hline
\multicolumn{1}{|c|}{}     & \multicolumn{1}{c|}{1}                & 0                & \multicolumn{1}{c|}{$+$}                            & \multicolumn{1}{c|}{$+$}       & $-$    & \multicolumn{1}{c|}{}                  & \multicolumn{1}{c|}{$\checkmark$} &              & \multicolumn{1}{c|}{}                  & \multicolumn{1}{c|}{$\checkmark$} &              \\ \cline{2-12} 
\multicolumn{1}{|c|}{$C_{1}$}    & \multicolumn{1}{c|}{0}              & 1                & \multicolumn{1}{c|}{$+$}                             & \multicolumn{1}{c|}{$+$}       & $-$    & \multicolumn{1}{c|}{}                  & \multicolumn{1}{c|}{}             &  & \multicolumn{1}{c|}{}      & \multicolumn{1}{c|}{} &   \\ \cline{2-12} 
\multicolumn{1}{|c|}{}     & \multicolumn{1}{c|}{1}                & 1.1                & \multicolumn{1}{c|}{$+$}                            & \multicolumn{1}{c|}{$+$}       & $-$   & \multicolumn{1}{c|}{$\checkmark$}      & \multicolumn{1}{c|}{}             &              & \multicolumn{1}{c|}{$\checkmark$}      & \multicolumn{1}{c|}{$\checkmark$} & $\checkmark$ \\ \hline
\end{tabular}
\caption{Relationship between the necessary parameters for the generation of squeezing synchronization and bipartite entanglement between bosonic modes.}
\label{t1}
\end{table*}

\subsubsection*{Case 1: Squeezing pump of the mechanical mode} 
Let us consider the mechanism of a squeezing driving of the mechanical oscillator mode, and analyze the stationary states of the bosonic modes in our system in the presence of the dissipation channels. Therefore, we numerically evaluate the influence of the tripartite interaction coupling, $\Lambda_{i}$, and the MO's squeezing pump strength, $q$, on the time evolution of the QF for all bosonic modes in the hybrid system.

For example, in Fig. \ref{fig5}$(a)$ one observes that in the absence of the the tripartite coupling for the second cavity, i.e. $\Lambda_{2}=0$, the second cavity being initially in a thermal state with very low excitation number ($\bar{n}_{c_2}=10^{-3}$), which can be approximated to almost vacuum state,  will maintain the value of vacuum fluctuations, i.e. $\langle(\Delta \mathcal{X}_{a_2})^{2}\rangle\approx 0.25$, throughout the dynamical process (see magenta line). In this case, squeezing is generated only in MO and first cavity (as $\Lambda_{1}=1$), see green and  blue lines, respectively. Therefore, a bipartite entanglement between the first cavity and MO, $\mathcal{N}(\hat{\rho}_{c_{1}},\hat{\rho}_{m})$, is realized while other kind of entanglement is absent, see Fig \ref{fig5}$(d)$.

In order to generate squeezing in the second cavity it is necessary to activate the tripartite interaction coupling, i.e. $\Lambda_{2}>0$ as appear in Fig. \ref{fig5} $(b,c)$. As consequence, the squeezing in the second cavity is generated while the squeezing in the first cavity is absent when $\Lambda_{1}=0$. In this case, a bipartite entanglement between the second cavity and MO, i.e. $\mathcal{N}(\hat{\rho}_{c_{2}},\hat{\rho}_{m})$ is present, as shown in Fig. \ref{fig8}$(e)$. On the other hand, when both tripartite interaction couplings coincide (i.e. $\Lambda_{1}/\omega_{m}=\Lambda_{2}/\omega_{m}=1$) one observes in Fig. \ref{fig5}$(c)$ the perfect synchronization between the modes of both cavities in the steady state. In this situation shown in Fig. \ref{fig5}$(f)$, where bipartite entanglement between cavity 1/cavity 2 and MO is equally distributed throughout the dynamics, while the bipartite entanglement between the cavities, quantified by $\mathcal{N}(\hat{\rho}_{c_{1}},\hat{\rho}_{c_{2}})$, reaches higher value, indicating so a strong stationary correlation between the cavities.

\subsubsection*{Case 2: Squeezing pump of the first cavity mode} 
As an alternative configuration, we induce squeezing in the first cavity by considering the pump Hamiltonian
\begin{equation}
    \mathcal{H}_{q'}=q'(a_{1}^{\dagger2}+a_{1}^{2}).
\end{equation}
As for the previous case, we vary the both tripartite interaction couplings, $\Lambda_{j}$, and find how all these mechanisms control the squeezing stabilization of the two cavities and MO modes. 
Similarly to case 1, in the absence of tripartite coupling ($\Lambda_{2}=0$ in this case) one observes that there is no squeezing for the second cavity mode, see Fig. \ref{fig6}$(a)$. Squeezing evolves only in the first cavity and MO. In addition, to generate squeezing synchronization between the MO and the cavities, it is necessary to choose the phase of QF, such that $\phi_{c_{1}}=\phi_{c_{2}}=\pi/4$ and $\phi_{m}=-\pi/4$. Therefore, the squeezed fields of the cavities and MO (on orthogonal axis)  can be efficiently synchronized in the steady-state. 

In general, the effect of squeezing evolution behaves similarly to the case 1, in fact to stimulate squeezing in the second cavity it is necessary to activate the tripartite interaction couplings, i.e. $\Lambda_{j}>0$. Therefore, one can reach squeezing synchronization between the MO and second cavity by controlling the tripartite interaction couplings and phase $\phi$ which appears in the definition of QF \ref{cuadrat222}, while unchanging the rest of parameters. Similarly to the case 1, the entanglement is distributed among all bosonic modes. However, in this case, the bipartite entanglement between the cavity 1(2) and MO is stronger than the entanglement between both cavities if comparing to the case 1, where the squeezing was pumped initially in the MO, see green curves in Figs. \ref{fig5}$(f)$ and \ref{fig6}$(f)$.

It is important to mention that for the two cases of the squeezing synchronization studied in this section, the tripartite entanglement is not generated in the hybrid network, as shown by the black curves in Figs. (\ref{fig5}-\ref{fig6}), and this conveys to a low fidelity  synchronization between the three bosonic modes. The complete results presented in this section are summarized in Tab. \ref{t1}.

\section{Discussion and outlook}\label{sec6}

\subsection{\textit{Effects of driving the three-level atoms and tripartite hybrid interaction}}

The role of the external driving of atoms and of the effective atom-photon-phonon interaction is important in the dynamical transfer of the quantum states between the cavities. In the following, we perform a qualitative description of how for example the squeezing is transferred from the first to the second cavity, considering the dynamical transitions. Let us define an arbitrary state as $|n_{a_1},n_{a_2},n_{c_{1}},n_{b},n_{c_{2}}\rangle$ where $n_{a_{1(2)}},n_{c_{1(2)}},n_{b}$ tag the excitation numbers in the first (second) atom, first (second) cavity and mechanical oscillator, respectively. Therefore, for an initial state $|00200\rangle$, where the first cavity is prepared in a squeezed state (i.e. a photon pair) and the rest subsystems are in ground states, then the Hamiltonian including the effective interaction and driving, $H_2+H_L$, can transfer the squeezed state from the first to the second cavity by the following dynamical procedure:

\begin{align}
\label{eq20}
    &|00200\rangle
    \xrightarrow[\Omega_{2}^{(1)}]{}
    |10200\rangle
    \xrightarrow[\sigma_{21,1}^{+}a_{1}b^{\dagger}]{}
    |20110\rangle
    \xrightarrow[\Omega_{1}^{(2)}]{}
    |22110\rangle
    \xrightarrow[]{}&\nonumber\\   
    &\xrightarrow[\sigma_{21,2}^{-}a_{2}^{\dagger}b]{}
    |21101\rangle
    \xrightarrow[\Omega_{1}^{(1)}]{}
    |01101\rangle
    \xrightarrow[\Omega_{2}^{(1)}]{}
    |11101\rangle
    \xrightarrow[]{}&\nonumber\\
   & \xrightarrow[\sigma_{21,1}^{+}a_{1}b^{\dagger}]{}
    |21011\rangle
    \xrightarrow[\Omega_{2}^{(2)}]{}
    |20011\rangle
    \xrightarrow[\Omega_{1}^{(2)}]{}
    |22011\rangle
    \xrightarrow[]{}&\nonumber\\
    &\xrightarrow[\sigma_{21,2}^{-}a_{2}^{\dagger}b]{}
    |21002\rangle.
\end{align}

\begin{figure*}[t]
\centering
\includegraphics[width=0.88\linewidth]{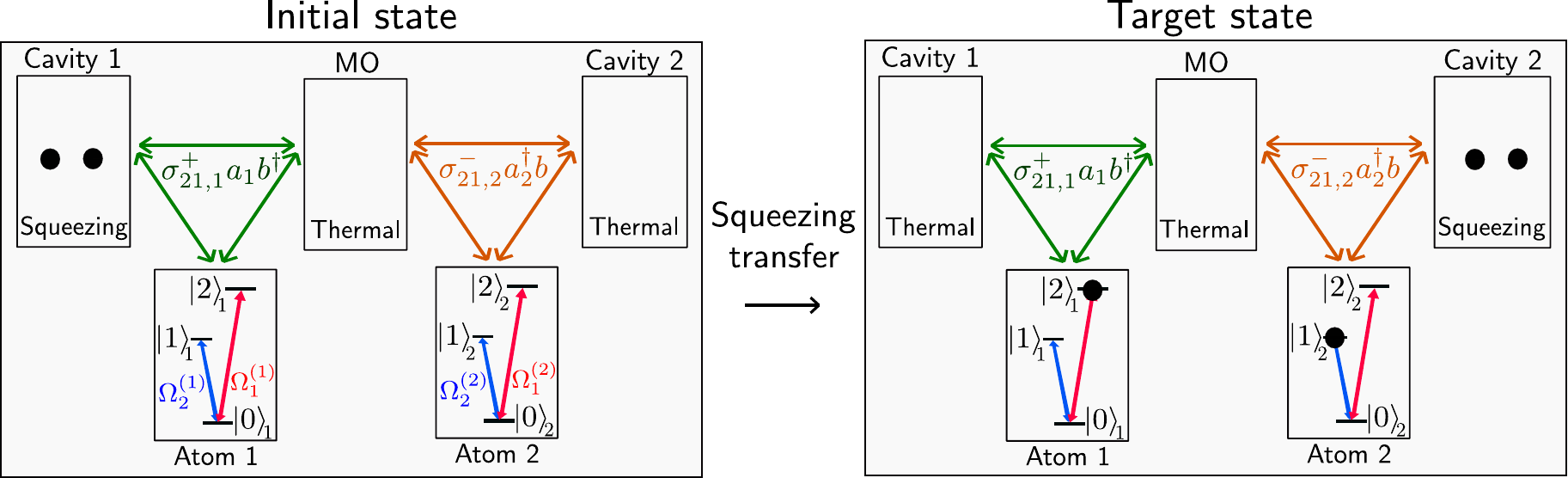}
\caption{Schematic illustration of the protocol discussed in Sec. \ref{sec1} which allows the transfer of squeezed state, i.e. a pair of excitations in cavity 1 (left panel) are finally transferred to cavity 2 (right panel), the evolution found in Figs.\ref{fig2}(c) and \ref{fig4}. Here, the process, $\sigma_{21, 1}^{+}a_{1}b^{\dagger}$, allows transfer of excitations to MO and the process, $\sigma_{21, 2}^{-}a_{2}^{\dagger}b$, allows transfer of excitations from MO to cavity 2. The activation of these tripartite interaction processes is clearly due to the result of driving of the three-level atoms to the necessary states, as represented in Eq.\ref{eq20} by the fields $\Omega_{1}^{(j)}$ and $\Omega_{2}^{(j)}$.}
\label{fig7}
\end{figure*}
In this probabilistic evolution, one finds that the driving process $\Omega_{2}^{(1)}$ of the atom in first cavity together with the effective tripartite coupling $\sigma_{21, 1}^{+}a_{1}b^{\dagger}$, allow the transfer of one excitation from the first cavity to MO, so the system evolves to the state $|20110\rangle$. Next, by driving the atom in the second cavity by $\Omega_{1}^{(2)}$ to its level $|2\rangle$, it is only possible to transfer the excitation from MO to the second cavity by the effective tripartite coupling $\sigma_{21, 2}^{-}a_{2}^{\dagger}b$, so ending in the state $|21101\rangle$. Up to this stage we have one excitation transferred from the first to the second cavity. Now, to transfer the second excitation from the first cavity, one should again create the excitation in MO, and so the driving processes with $\Omega_{1}^{(1)}$ and $\Omega_{2}^{(1)}$ allow the possibility of transferring this excitation to the MO by using the same tripartite coupling $\sigma_{21,1}^{+}a_{1}b^{\dagger}$, ending in the state $|21011\rangle$. Finally, the transfer of the MO's excitation to the second cavity is only possible by the process $\sigma_{21,2}^{-}a^{\dagger}_{2}b$ and for this, the atom in the second cavity must be prepared in the level $|2\rangle$ through the level $|1\rangle$, respectively by the driving processes with $\Omega_{2}^{(2)}$ and $\Omega_{1}^{(2)}$, see Fig. \ref{fig1}(b). As result, the system ends up in the state $|21002\rangle$, so transferring the squeezed state from the first to the second cavity, as observed in Fig.\ref{fig2}(c) at the time $t_2$. The protocol is sketched in Fig.\ref{fig7}.

\subsection{\textit{Experimental feasibility}}

In our numerical calculations we have considered the optomechanical coupling within a range $\lambda \leq 0.01\omega_m$, and if taking for example $\omega_m/(2\pi)\approx 2$ GHz to have low number of thermal excitations (as explained in Sec.\ref{sec3}c), so we get the top limit value $\lambda/(2\pi)\approx 20$ MHz. This is considered a strong optomechanical coupling, however we hope is experimentally feasible, particularly considering very recent experiment \cite{Enzian19}, where the optomechanical coupling $ \propto 40$ MHz was reached. Alternatively, the optomechanical coupling can be increased in setups such as those discussed in \cite{Pirk2015,Monte2022,Leij2015}. For the squeezing synchronization protocol we considered strong dissipation situation, with the cavity damping rate as $\kappa_a/(2\pi)= 400$ MHz and MO damping rate as $\kappa_b/(2\pi)=4$ MHz (see Fig.\ref{fig4}), which are compatible with the recent experiments in the optomechanics \cite{Aspelmeyer,Mir2020,Rie2016}.

\section{Conclusion}\label{sec7}

We have proposed a hybrid optomechanical network consisting of two-mode cavity, equivalent to two cavities with a three-level atom in each one, and the movable mirror (mechanical oscillator) allows the transfer of quantum states between the cavities. In our setup it is also possible to generate a bipartite entanglement between the bosonic modes and prepare stationary squeezed states of mechanical and cavity modes. We find that when two external fields independently driving each atom, the squeezed and Schr\"{o}dinger's cat states between the cavities can be transferred with an extremely high fidelity under the unitary dynamics, see Figs. \ref{fig2},\ref{fig8}. In this framework we are able to show the dynamical generation and distribution of bipartite and tripartite entanglement, see Fig. \ref{fig4}. As the hybrid system evolves in time, then the cavity-cavity transfer protocol is weakened by the fact that the tripartite interaction between the bosonic modes (two photonic and phonon) becomes stronger and so the tripartite entanglement increases, see black curve in Fig. \ref{fig4}. At the instants where the transfer of quantum state occurs, the bipartite and tripartite entanglement almost vanish. 

Additionally, in case of highly dissipative dynamics of the hybrid optomechanical system with the driving atoms and using coherent pumping of squeezed phonons/photons in the initial stage, one can synchronize a pair of bosonic modes in squeezed steady states for the bipartite systems as cavity 1-cavity 2, cavity1-MO and cavity 2-MO, see Figs. \ref{fig5} and \ref{fig6}. The effect of squeezing synchronization of the cavities and mechanical modes can be achieved regardless of the pump mechanism discussed here. This result facilitates the experimental performance by choosing an opportune pumping mechanism, find more details in Table \ref{t1}.

We hope that this study will push forward the development of optomechanical networks, where quantum protocols such as the generation, transfer and stabilization of quantum states and correlations are essential. For example, there are many sensing applications where setups need stabilized and transient squeezed states.


\section*{acknowledgments}
We thank Prof. Miguel Orszag for his contribution in the initial stage of this research. H.M. acknowledge Universidad Mayor through the Doctoral scholarship during which the work began. B.H and V.E. acknowledge the financial support from ANID Fondecyt Regular No. $1221250$. V.E. acknowledge grant No. $20.80009.5007.01$ of the State Program (2020-2023) from National Agency for Research and Development of Moldova. 

\appendix
\section{\label{apendice1}Derivation of the effective Hamiltonian}

Here we show how to transform the Hamiltonian in Eq. \ref{base} to the interaction picture. The first transformation is: 
\begin{align}\label{int}
    \mathcal{V}&=e^{\imath\mathcal{H}_{0} t}\mathcal{H}_{I}e^{-\imath\mathcal{H}_{0} t},
\end{align}
where
\begin{align}
\mathcal{H}_{0}&=\omega_{m}b^{\dagger}b+\sum_{j=1}^{2}\sum_{i=0}^{2}\omega_{i,j}\sigma_{ii,j}+\omega_{c_{j}}a_{j}^{\dagger}a_{j},\\
    \mathcal{H}_{I}&=\sum_{j=1}^{2}g_{j}\left(a_{j}\sigma^{+}_{21,j}+a_{j}^{\dagger}\sigma^{-}_{21,j}\right) \nonumber \\
    &- \lambda\left(a_{1}^{\dagger}a_{1}-a_{2}^{\dagger}a_{2}\right)\left(b+b^{\dagger}\right).
\end{align}
From Eq. \ref{int} we get
\begin{eqnarray}
    \mathcal{V}&=& \sum_{j=1}^{2}g_{j}\left(a_{j}\sigma^{+}_{21,j}e^{\imath\Delta_{j}t}+a_{j}^{\dagger}\sigma^{-}_{21,j}e^{-\imath\Delta_{j}t}\right) \nonumber \\
    &-&\lambda\left(a_{1}^{\dagger}a_{1}-a_{2}^{\dagger}a_{2}\right)\left(be^{\imath\omega_{m}t}+b^{\dagger}e^{-\imath\omega_{m}t}\right),
\end{eqnarray}
 where $\Delta_{j}=\omega_{2,j}-\omega_{1,j}-\omega_{c_{j}}$.
 
Now, we move to a second interaction picture
\begin{equation} \label{int2}
    \mathcal{V}'=\exp{\left\{\imath\int\mathcal{V}_{0}dt\right\}}\mathcal{V}_{I}\exp{\left\{-\imath\int\mathcal{V}_{0}dt\right\}},
\end{equation}
where
\begin{align}
    \mathcal{V}_{0}&=\left(a_{2}^{\dagger}a_{2}-a_{1}^{\dagger}a_{1}\right)f(t),\\
    \mathcal{V}_{I}&=\sum_{j=1,2}g_{j}\left(a_{j}\sigma^{+}_{21,j}e^{\imath\Delta_{j}t}+a_{j}^{\dagger}\sigma^{-}_{21,j}e^{-\imath\Delta_{j}t}\right),\\
    f(t)&=\lambda\left(be^{\imath\omega_{m}t}+b^{\dagger}e^{-\imath\omega_{m}t}\right).
\end{align}
As result of Eq. \ref{int2} one gets
\begin{eqnarray}\label{1eq}
\mathcal{V}'=\sum_{j=1,2}g_{j}a_{j}\sigma^{+}_{21,j}e^{\imath\left[\Delta_{j}t+(-1)^{j+1}F(t)\right]}+H.c.,
\end{eqnarray}
where
\begin{equation}
    F(t)=\int f(t)dt=\frac{\lambda}{\imath\omega_{m}}\left(b^{\dagger}\eta-b\eta^{*}\right),
\end{equation}
with $\eta=e^{\imath\omega_{m}t}-1$.


\begin{figure*}[t]
\centering
\includegraphics[width=0.45\linewidth]{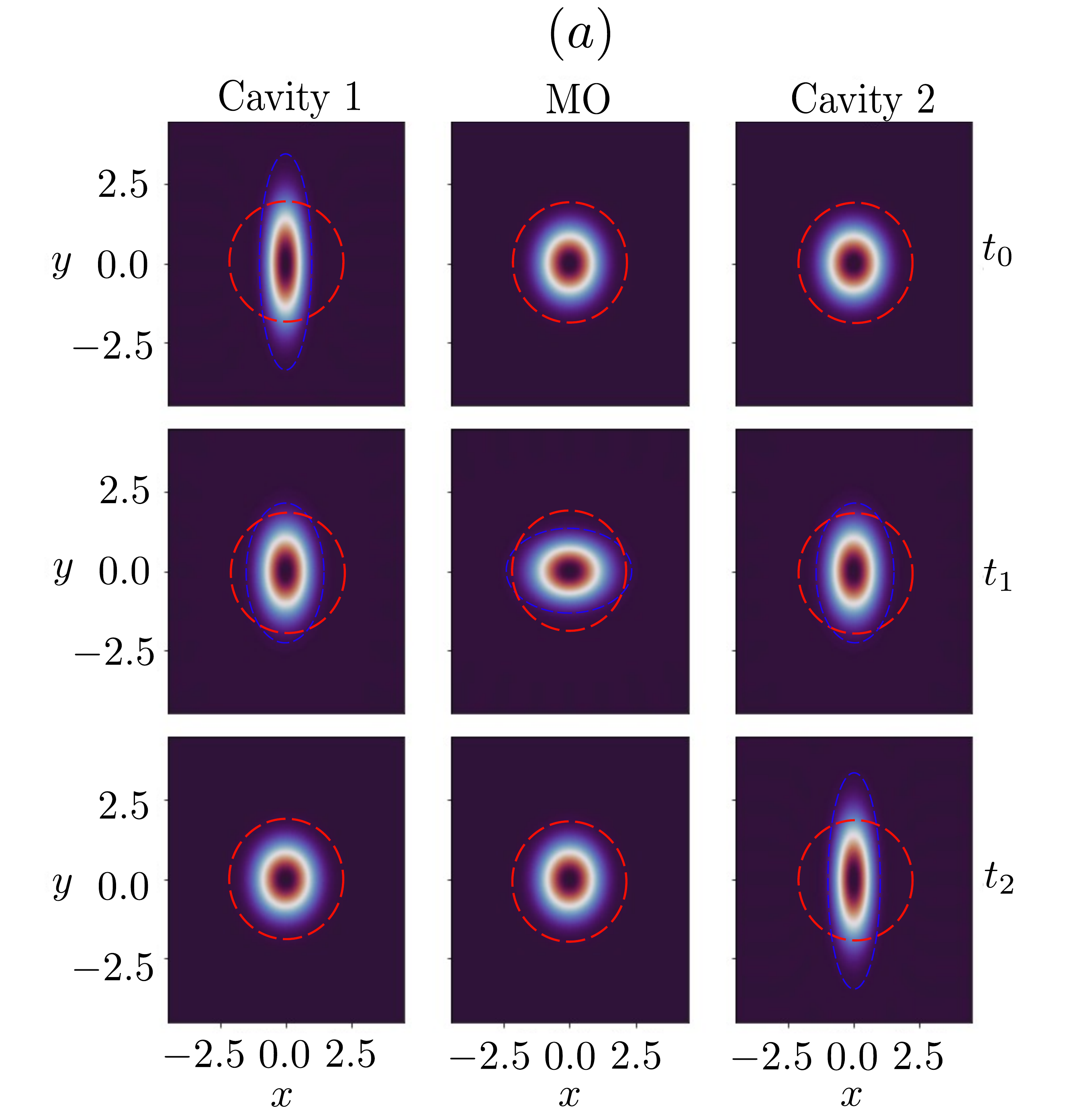}
\hspace{0.4cm}
\includegraphics[width=0.45\linewidth]{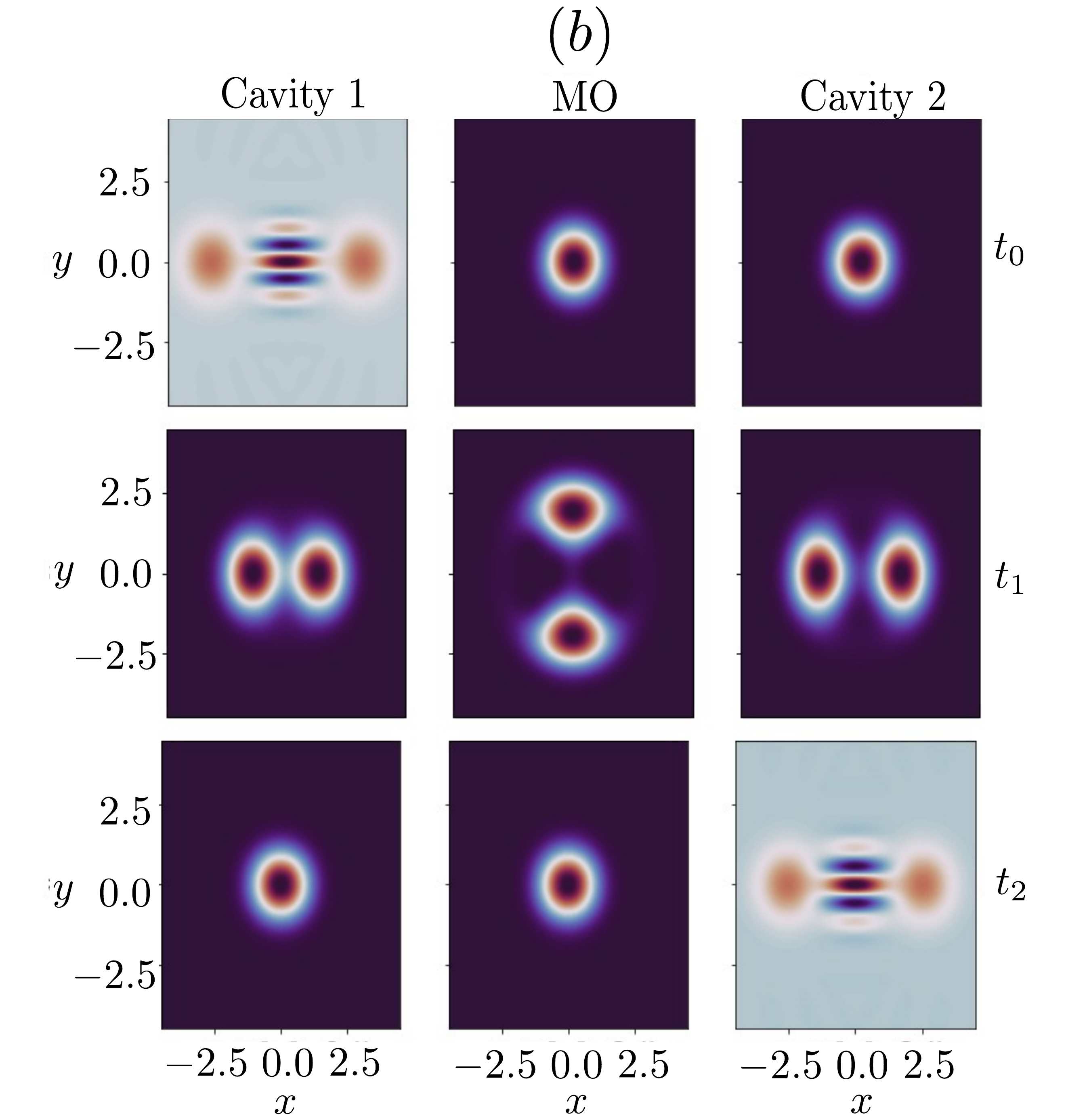}
\caption{Wigner quasi-probability distribution for the states of cavities and MO at different time instants. One observes the effect of squeezed (left) and cat (right) state transfer between the cavities. The parameters are the same as in Fig. \ref{fig2}.}
\label{fig8}
\end{figure*}

\textit{Approximation:}
In our model we assume the optomechanical coupling $\lambda$ is much smaller than the mechanical frequency $\omega_{m}$, so that
$e^{-\imath(-1)^{j} F(t)}\approx1-\imath(-1)^{j}\lambda\left(b^{\dagger}\eta-b\eta^{*}\right)/\omega_{m}$.
So Eq. \ref{1eq} takes the form

\begin{eqnarray} \label{eqA11}
    \mathcal{V}'
    &=&\sum_{j=1,2} g_{j}a_{j}\sigma^{+}_{21,j}e^{\imath\Delta_{j}t} +(-1)^{j}\varLambda_{j} \bigg [ a_{j}\sigma^{+}_{21,j}\big(b^{\dagger}-b\big)
    e^{i\Delta_{j}t} \nonumber \\
    &-& a_{j}\sigma^{+}_{21,j}\bigg(b^{\dagger} 
    e^{\imath(\Delta_{j}+\omega_{m})t} -b e^{\imath(\Delta_{j}-\omega_{m})t}\bigg) \bigg] + H.c.
\end{eqnarray}
where $\varLambda_{j}=g_{j}\cdot\lambda/\omega_{m}$. Now, by considering the blue-detuned regime $\Delta_{j}=-\omega_{m}$ we get
\begin{eqnarray} \label{eqA12}  
\mathcal{V}'
    &=\sum_{j=1,2} g_{j}a_{j}\sigma^{+}_{21,j}e^{-\imath \omega_{m}t} + (-1)^{j}\varLambda_{j} \bigg [ a_{j}\sigma^{+}_{21,j}\big(b^{\dagger}-b\big)
    e^{-i\omega_{m} t}  \nonumber \\
    &- a_{j}\sigma^{+}_{21,j}\bigg(b^{\dagger} -b e^{-2 \imath \omega_{m} t}\bigg) \bigg] + H.c.,
\end{eqnarray}
Neglecting time-dependent terms, we obtain the effective tripartite interaction
\begin{eqnarray} \label{eqA13}
    \mathcal{V}'=\sum_{j=1,2}(-1)^{j+1}\varLambda_{j} a_{j}\sigma^{+}_{21,j}b^{\dagger}+H.c.
\end{eqnarray}

\section{\label{apendice2}Wigner visualization}
For a better visualization of the results shown in Fig. \ref{fig2}, we present here the Wigner quasi-probability distribution in three different instants, i.e. $\left\{t_0, t_1, t_2\right\}$, of the dynamic evolution in Fig.\ref{fig2}. It can be easily detected the squeezed and Schr\"{o}dinger's cat states in the initial time ($t_0$) and transitory evolution time ($t_2$), where the target state is transferred.

In Fig. \ref{fig8}$a$ we see how the initial squeezed state (blue dashed ellipse in the top panel) in cavity 1 disappears (central panel) and tends to form a thermal state (red dashed circle) while cavity 2, initially in thermal state begins to squeeze (blue dashed ellipse). Finally in the lower panel of $(a)$ one finds the squeezed state in cavity 2, which looks similar to the initial state of the cavity 1. While cavity 1 and MO end in thermal states.
In Fig. \ref{fig8}$b$ we show an equivalent transfer effect for a Schr\"{o}dinger's cat state, in which is initialized the mode in cavity 1. 
\newpage

\bibliography{apssamp}

\end{document}